%% file: draft.tex
\documentclass[a4paper,11pt]{article}
\pdfoutput=1
\usepackage{jheppub}

\usepackage[utf8]{inputenc}
\usepackage[T1]{fontenc}
\usepackage{lmodern}
\usepackage{textcomp}
\usepackage{microtype}

\usepackage{units}
\usepackage[font=small]{caption}
\usepackage[font=footnotesize]{subcaption}
\newcommand*{\doi}[1]{\href{http://dx.doi.org/#1}{doi: #1}}

\usepackage{booktabs}

\usepackage{multirow}

\usepackage{tikz}
\usetikzlibrary{positioning}
\usetikzlibrary{fit}
\usetikzlibrary{external}
\newcommand{\Br}{\ensuremath{\text{Br}}}

\AtBeginDocument{%
\mathchardef\mathcomma\mathcode`\,
\mathcode`\,="8000}
{\catcode`,=\active
\gdef,{\mathcomma\discretionary{}{}{}}}

\usepackage{xspace}
\newcommand{\software}[2][]{\texttt{#2}\xspace#1}

\title{Heavy Higgs Bosons at \unit[14]{TeV} and \unit[100]{TeV}}

\author[a,b]{Jan Hajer,}
\author[a]{Ying-Ying Li,}
\author[a]{Tao Liu,}
\author[a]{John F.H. Shiu}

\emailAdd{jan.hajer@ust.hk}
\emailAdd{ylict@connect.ust.hk}
\emailAdd{taoliu@ust.hk}
\emailAdd{fhshiu@ust.hk}

\affiliation[a]{Department of Physics, The Hong Kong University of Science and Technology,
\\ Clear Water Bay, Kowloon, Hong Kong S.A.R., P.R.C.}
\affiliation[b]{Jockey Club Institute for Advanced Study, The Hong Kong University of Science and Technology,
\\ Clear Water Bay, Kowloon, Hong Kong S.A.R., P.R.C.}

\abstract{%
Searching for Higgs bosons beyond the Standard Model (BSM) is one of the most important missions for hadron colliders.
As a landmark of BSM physics, the MSSM Higgs sector at the LHC is expected to be tested up to the scale of the decoupling limit of \unit[$\mathcal O(1)$]{TeV}, except for a wedge region centered around $\tan\beta \sim 3 -10$, which has been known to be difficult to probe.
In this article, we present a dedicated study testing the decoupled MSSM Higgs sector, at the LHC and a next-generation $pp$-collider, proposing to search in channels with associated Higgs productions, with the neutral and charged Higgs further decaying into $tt$ and $tb$, respectively.
In the case of neutral Higgs we are able to probe for the so far uncovered wedge region via $pp\to bb H/A \to bbtt$.
Additionally, we cover the the high $\tan\beta$ range with $pp\to bb H/A \to bb\tau\tau$.
The combination of these searches with channels dedicated to the low $\tan\beta$ region, such as $pp\to H/A \to tt$ and $pp\to tt H/A \to tttt$ potentially covers the full $\tan\beta$ range.
The search for charged Higgs has a slightly smaller sensitivity for the moderate $\tan\beta$ region, but additionally probes for the higher and lower $\tan\beta$ regions with even greater sensitivity, via $pp\to tb H^\pm \to tbtb$.
While the LHC will be able to probe the whole $\tan\beta$ range for Higgs masses of $\mathcal O(1)$ TeV by combining these channels, we show that a future \unit[100]{TeV} $pp$-collider has a potential to push the sensitivity reach up to $\sim \mathcal O(10)$ TeV.
In order to deal with the novel kinematics of top quarks produced by heavy Higgs decays, the multivariate Boosted Decision Tree (BDT) method is applied in our collider analyses.
The BDT-based tagging efficiencies of both hadronic and leptonic top-jets, and their mutual fake rates as well as the faking rates by other jets ($h$, $Z$, $W$, $b$, etc.) are also presented.

\vspace*{5mm} \noindent \today
}

\begin{document}

\maketitle

\section{Introduction}
\label{sec:int}

With the discovery of a \unit[125]{GeV} Higgs boson and rapid progress made at the Large Hadron Collider (LHC), it is time for the high-energy physics community to chart a road-map for the next decade or the next  few decades.
In addition to a high-luminosity LHC program, two preliminary proposals, the Future hadron-hadron Circular Collider (Fcc$_{\rm hh}$) program at CERN~\cite{Gomez:2013zzn} and the Super-$pp$-Collider (SppC)~\cite{CEPC-SppC} program in China, have been made, both of which involve construction of a \unit[50--100]{TeV} $pp$ collider~\cite{Hinchliffe:2015qma} (below we will universally assume a \unit[100]{TeV} machine).
With more accumulated data or higher expected energy scale, the high-luminosity LHC and the next-generation $pp$-colliders offer great opportunities for the search for physics up to and beyond TeV scale, respectively, including additional Higgs bosons.
(For recent studies on physics at a next-generation $pp$-collider, e.g, see~\cite{Curtin:2014jma, Bramante:2014tba, Auerbach:2014xua, Butterworth:2014efa, He:2014xla, Avetisyan:2013onh, Avetisyan:2013dta, Anderson:2013kxz, Assadi:2014nea, Ward:2014ksa, Hook:2014rka, Carena:2014nza})

An extended Higgs sector extensively exists in physics beyond the Standard Model (BSM), such as supersymmetric theories~\cite{Nilles:1992una} or composite Higgs models~\cite{Georgi:1984af}, because of requirements by either symmetry or phenomenology. The additional Higgs fields fill up singlet, doublet, triplet or some other representations of electroweak (EW) gauge symmetry, generically yielding interactions with the SM sector which are characterized by their electric charges and CP-structures. Searching for these new Higgs bosons, therefore, provides an unambiguous way to probe for new physics and is one of the top missions of hadron colliders.

At the LHC, both the ATLAS and the CMS  collaborations have started their searches for neutral, singly-charged or doubly-charged Higgs bosons, in various channels~\cite{Aad:2014kga, Khachatryan:2014wca, Aad:2014vgg, CMS-PAS-HIG-14-020, Chatrchyan:2012ya, ATLAS:2012hi}.
Within the next decade, tests up to $\unit[\mathcal O(1)]{TeV}$ are  expected in a general context.
Additionally, the energy scale accessible to the next-generation $pp$-collider is several times higher than that of the LHC, which enables us to search for a much heavier Higgs sector. In both cases, some decay modes kinematically suppressed in the low mass domain are fully switched on.
Concurrently, the kinematics of their decay products can be dramatically changed.
A systematic study of decoupled or heavy Higgs sectors at hadron colliders is therefore essential for both the high-luminosity LHC program and the proposals for next-generation $pp$-colliders.

This motivates the studies in this article.
We focus on searches for neutral and singly-charged heavy Higgs bosons, and will analyze the sensitivity reach that might be achieved at both, the (HL-)LHC and a \unit[100]{TeV} $pp$ collider, using the Minimal Supersymmetric Standard Model (MSSM) for illustration.
In the MSSM, there are in  total five Higgs bosons: three of them are neutral (two CP-even and one CP-odd, or three CP-mixing ones) and two of them are charged.
Because of the limitation in the energy reach, the LHC searches mainly focus on a mass domain below the TeV scale.
For constraints on the MSSM Higgs sector based on these searches see e.g. \cite{Djouadi:2015jea, Bhattacherjee:2015sga}.
For searches including light gauginos and higgsinos see for example \cite{Arganda:2013ve,Arganda:2012qp}.

For neutral Higgs bosons, $pp \to bb H/A \to bb \tau\tau$ and $pp \to H \to VV,\ hh,\ tt$ together with $pp \to A \to hZ,\ tt$ yield or might yield the best sensitivities in the large $\tan\beta$ region ($\tan\beta > 10$) as well as the small $\tan\beta$ region ($\tan\beta < 3$), respectively.
Whereas the moderate $\tan\beta$ region ($\tan\beta \sim 3-10$) is difficult to probe, yielding a well-known untouched ``wedge'' region (for recent discussions, e.g., see~\cite{Djouadi:2015jea, Gennai:2007ys}).
As for charged Higgs bosons, $pp \to t b H^\pm \to tb \tau \nu$ plays a crucial role in probing the small $\tan\beta$ region with $m_{H^\pm} < m_t + m_b$ as well as the large $\tan\beta$ region, because the coupling of $H^\pm$ with $\tau \nu$ is $\tan\beta$-enhanced.

\begin{table}
\centering
\begin{tabular}{l|l|l} \hline
  & $\tan\beta$
  & Channels
 \\ \hline
  & High
  & $pp\to bb H/A \to bb \tau\tau, bbbb^*$
 \\ \cline{2-3}
    Neutral Higgs ($H/A$)
  & Moderate
  & $pp \to bb H/A \to bb tt$
 \\ \cline{2-3}
  & Low
  & $pp \to H/A \to tt^*$, $pp\to ttH/A \to tttt^*$
 \\ \hline
    \multirow{2}{*}{Charged Higgs ($H^\pm$)}
  & High
  & $pp\to tb H^\pm \to tbtb, tb \tau \nu$
 \\ \cline{2-3}
  & Low  & $pp\to tb H^\pm \to tb tb$
 \\ \hline
\end{tabular}
\caption{Main channels to cover or potentially cover various $\tan\beta$ regions in the decoupling limit of the MSSM Higgs sector. The channels marked by ``$*$'' are not covered in collider analyses in this article.}
\label{tab:decay_tra}
\end{table}

In the decoupling limit, the decays of $H/A \to tt$ and $H^\pm \to tb$ are fully switched on, if the SUSY sector is decoupled as well, whereas the decays of $H/A \to VV,\ hh$, $A \to hZ$, and $H^\pm \to h W^\pm$ are generically suppresse~\cite{Djouadi:2005gj}.
$\Br(H/A \to tt)$ becomes sizable for moderate $\tan\beta$ and dominant for low $\tan\beta$, and $\Br(H^\pm \to tb)$ becomes dominant for the whole $\tan\beta$ region.
However, it is known that the signal in the channel $pp \to H/A \to tt$ and the QCD $tt$ background have strong interference effects~\cite{Frederix:2007gi, Dicus:1994bm}.
The search in this channel therefore is extremely challenging.
Instead we propose in this article to test a heavy Higgs sector in channels with associated Higgs productions, where the interference effects between the signal and the QCD background are much less severe, compared to that in $pp\to H/A \to tt$.

Interestingly, the channels $pp\to bb H/A \to bb tt$ and  $pp\to tt H/A \to tt tt$ have a cross section maximized in the moderate $\tan\beta$ and low $\tan\beta$ regions, respectively.  A combination of the channels of $pp\to bb H/A \to bb \tau\tau$ (or $pp\to b H/A \to bbb$~\cite{Carena:2012rw}),  $pp\to bb H/A \to bb tt$ and $pp \to ttH/A \to tttt$ or $pp \to H/A \to tt$ (if the interference structure can be efficiently identified) thus may yield a full coverage for the  $\tan\beta$ domain in searching for neutral Higgs bosons, if the moderate and low $\tan\beta$ regions can be probed in the two latter channels efficiently. For the charged Higgs boson searches $pp \to tb H^\pm \to tbtb$ is the golden channel for both high and low $\tan\beta$ domains.

The main channels for the MSSM Higgs searches, which will be explored in this paper (except the ones marked with ``$*$''), are summarized in Table~\ref{tab:decay_tra}. These channels are characterized by two classes of kinematics:
\begin{enumerate}
\item Kinematics related to the heaviness of the BSM Higgs bosons, such as highly boosted top quarks in Higgs decay. Looking into internal structure of the boosted objects or defining boostness-based variables can efficiently suppress the related backgrounds.

\item Kinematics related to the particles accompanying Higgs production, e.g., the forwardness/backwardness of the two $b$-jets accompanying the Higgs production in the channels of $pp\to bbH/A \to bbtt$ and $pp \to tb H^\pm\to tbtb$, and the reconstructed top pair accompanying the Higgs production in $pp\to tt H/A \to tttt$.
\end{enumerate}
Meanwhile, the kinematic features of the particles accompanying Higgs production can efficiently facilitate revealing the $m_{tt}$ or $m_{tb}$ peak of the signals.

To fully extract the potential of a $pp$ collider, particularly one at \unit[100]{TeV}, in searching for the decoupled MSSM Higgs sector, we exploit the kinematic features discussed above by applying the multivariate Boost-Decision-Tree (BDT) method for the channels involving top quarks.
As a demonstration of the effectiveness of the BDT method, we present, additionally to the sensitivity reach of the $pp$ colliders at \unit[14 and 100]{TeV}, the tagging efficiencies of both hadronic and leptonic top-jets, and their mutual fake rates as well as the faking rates by other jets ($h$, $Z$, $W$, $b$, etc.). The code implementing our analyses \software[0.1]{BoCA} is publicly available \cite{Boca}.
Although the studies focus on the MSSM Higgs sector for its theoretical motivation and experimental representativeness, the sensitivity reach can be projected straightforwardly to some other contexts such as  two Higgs doublet models and composite Higgs models. The strategies developed for these Higgs searches might be applied to the other collider analyses as well.

We organize this article in the following way.
We shortly review the MSSM Higgs sector in the decoupling limit in Section 2, and introduce the strategies for constructing the BDT-based top-jet taggers (both hadronic and leptonic) and testing the decoupled MSSM Higgs sector at $pp$ colliders in Section 3. The discovery reaches and exclusion limits which might be achieved at \unit[14]{TeV} and \unit[100]{TeV} are presented in Section~4.
We summarize our studies and point out potential directions to exploration after this work in Section~5.
More details on the collider analyses and the construction of the BDT-based top-taggers are provided in the Appendices.

\section{The Higgs Sector in the MSSM}
\label{sec:the}

\begin{table}
\centering
\begin{tabular}{l|c|c }
\hline
  & Couplings
  &  MSSM
 \\ \hline \multirow{4}[4]{*}{$H$}
  & $g_{H VV}$
  & $ \cos (\beta-\alpha)$
 \\
  & $g_{H t \bar t}$
  & $ \sin\alpha/\sin\beta$
 \\
  & $g_{H b \bar b}$
  & $ \cos\alpha/\cos\beta$
 \\
  & $g_{H \tau \bar \tau}$
  & $\cos\alpha/\cos\beta$
 \\ \hline \multirow{4}[4]{*}{$A$}
  & $g_{A VV}$
  & 0
 \\
  & $g_{A t \bar t}$
  & $ \cot \beta$
 \\
  & $g_{A b \bar b}$
  & $ \tan \beta$
 \\
  & $g_{A \tau \bar \tau}$
  & $ \tan \beta$
 \\ \hline \multirow{4}[6]{*}{$H^\pm$}
  & $g_{H^+ \bar u d}$
  & $ \frac{1}{ \sqrt{2} v}  V_{ud}^* [m_d \tan \beta (1+\gamma_5) + m_u{\rm cot}\beta (1-\gamma_5)] $
 \\
  & $g_{H^- u \bar d}$
  & $\frac{1}{ \sqrt{2} v}  V_{ud} [m_d \tan \beta (1-\gamma_5) + m_u{\rm cot}\beta (1+\gamma_5)]  $
 \\
  & $g_{H^+ \bar \nu l}$
  & $\frac{1}{ \sqrt{2} v}  m_l \tan \beta (1+\gamma_5)$
 \\
  & $g_{H^- \nu \bar l}$
  & $\frac{1}{ \sqrt{2} v}  m_l \tan \beta (1-\gamma_5) $
 \\ \hline
\end{tabular}
\caption{Tree-level $H/A$ and $H^\pm$ couplings in the MSSM. Here $V_{ud}$ is the CKM matrix element present in the case of quarks.}
\label{tab:2HDMcoupling}
\end{table}

In the MSSM, the Higgs mass spectrum and their couplings with the SM sector and themselves depend, at tree level, only on two free parameters, often chosen to be the Higgs vacuum expectation value (VEV) alignment $\tan\beta = \frac{v_2}{v_1}$ and the mass of the CP-odd neutral Higgs boson $m_A$ (or the mass of the charged Higgs boson $m_{H^\pm}$) in the case with no CP-violation~\cite{Djouadi:2005gj}.
In comparison, the general type II 2HDM depends on more parameters at tree level~\cite{Gunion:1989we, Ginzburg:2002wt}.
The additional free parameters include the masses of the three other Higgs bosons, and the mixing angle $\alpha$ defined within the CP-even neutral sector by
\begin{equation}
\label{eq:hH2HDM}
\begin{pmatrix}
h \\ H
\end{pmatrix}
=
\begin{pmatrix}
- \sin \alpha & \cos \alpha \\
\cos \alpha & \sin \alpha
\end{pmatrix}
\begin{pmatrix}
H^0_1 \\ H^0_2
\end{pmatrix}
\end{equation}
if CP-symmetry is conserved~\cite{Ginzburg:2002wt}.
In the MSSM these parameters are correlated to each other.
For example, the mixing angle $\alpha$
is fixed by $\tan\beta$ and $m_A$ with the relation (e.g. see~\cite{Martin:1997ns})
\begin{align}
    \alpha
  = \frac{1}{2} \arctan \left( \tan 2 \beta \, \frac{m_A^2 + M_Z^2}{ m_A^2 - M_Z^2} \right) \ ,
  && - \frac{\pi}{2} \leq \alpha \leq 0 \ .
\label{alpha:tree}
\end{align}
In addition to the SM sector and with themselves, the MSSM Higgs bosons can couple with superparticles directly, which may significantly alter their productions and decays.  Instead of giving a full phenomenological consideration, we assume a decoupled SUSY sector~\cite{Haber:1995be}. We neglect supersymmetric corrections to conventional Higgs productions and decays, tolerate a potential deviation of the SM-like Higgs mass from the observed value, and turn off the Higgs decays into superparticles. Such a treatment ensures the analyses be less model-dependent, and enables us project the sensitivity reach to a specific model more easily.

\begin{figure}
\begin{subfigure}{.32\textwidth}
\includegraphics[width=\textwidth]{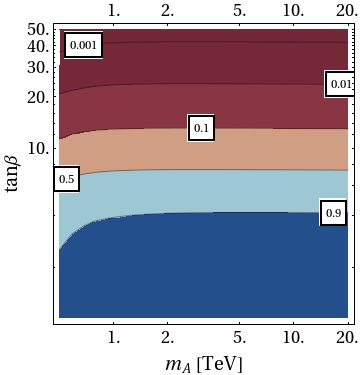}
\caption{$\Br(H\to tt)$}
\label{fig:BrCPEven}
\end{subfigure}
\hfill
\begin{subfigure}{.32\textwidth}
\includegraphics[width=\textwidth]{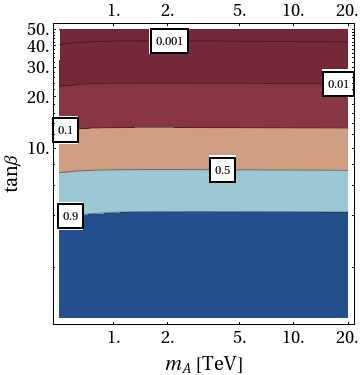}
\caption{$\Br(A\to tt)$}
\label{fig:BrCpOdd}
\end{subfigure}
\hfill
\begin{subfigure}{.32\textwidth}
\includegraphics[width=\textwidth]{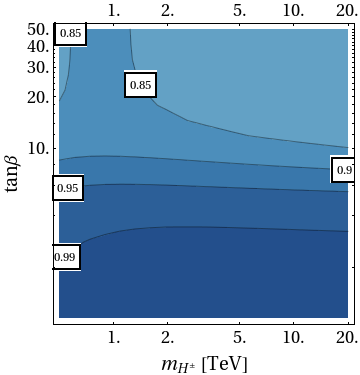}
\caption{$\Br(H^\pm \to tb)$}
\label{fig:BrCharged}
\end{subfigure}
\caption{Contours for the branching ratios of CP-even neutral Higgs to top pairs~\subref{fig:BrCPEven}, CP-odd neutral Higgs to top pairs~\subref{fig:BrCpOdd} and charged Higgs to a top-bottom pair~\subref{fig:BrCharged} in the MSSM.
}
\label{fig:brcont}
\end{figure}

In the decoupling limit, the couplings take the values of the 2HDM alignment limit
\begin{eqnarray}
    g_{HVV} = g_{hZA} = g_{hW^\mp H^\pm}  \propto \cos(\beta-\alpha)  \to 0
\end{eqnarray}
yielding a generic suppression in sensitivities for the search modes (see Table~\ref{tab:2HDMcoupling}): $H \to VV$, $A \to hZ$, and $H^\pm \to h W^\pm$.
As for $H\to hh$, though the involved coupling $g_{Hhh}$ is not completely suppressed in the alignment limit, its decay width is inversely proportional to $m_H$, yielding a suppressed sensitivity for this mode in the case of large $m_H$ (e.g., see~\cite{Djouadi:2015jea}).
In the decoupling limit, therefore, $H/A$ mainly decay into $bb,\ \tau\tau$ for large and intermediate $\tan\beta$ and into $tt$ for intermediate and low $\tan\beta$.
$H^\pm$ overwhelmingly decay into $tb$ except in the parameter region with relatively small $m_{H^\pm}$ and large $\tan\beta$ where the branching ratio of $H^\pm$ into $\tau\nu$ is not negligibly small.
The total decay width of these Higgs bosons varies at percent level or even below, scaled by their mass.

The contours for branching fractions of $H/A \to tt$ and $H^\pm \to tb$ are  shown in Figure~\ref{fig:brcont}.
The figure indicates a clear $\tan\beta$ suppression for these branching fractions.
Compared to $\Br(H/A \to tt)$, however, $\Br(H^\pm \to tb)$ is much less suppressed in moderate and large $\tan\beta$ regions.
This is mainly because the coupling $g_{H^\pm \to tb}$ is at tree level a linear combination of the $t$ and $b$ Yukawa couplings, with the latter yileding a $\tan\beta$-enhanced contribution to the $H^\pm$ decay width. In spite of this,
$\Br(H^\pm \to tb)$ is relatively small in the upper-left corner, due to additional phase-space suppression.

\begin{figure}
\begin{subfigure}{.32\textwidth}
\includegraphics[width=\textwidth]{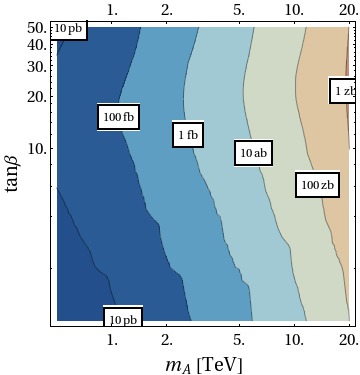}
\caption{$\sigma(pp \to H/A)$}
\label{fig:ProdHA}
\end{subfigure}
\hfill
\begin{subfigure}{.32\textwidth}
\includegraphics[width=\textwidth]{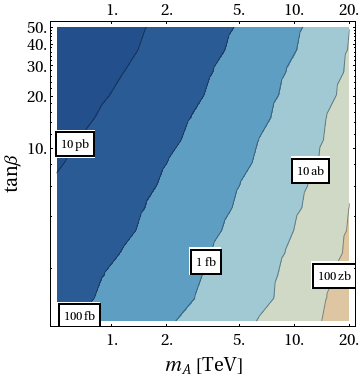}
\caption{$\sigma(pp \to bb H/A)$}
\label{fig:ProdHAbb}
\end{subfigure}
\hfill
\begin{subfigure}{.32\textwidth}
\includegraphics[width=\textwidth]{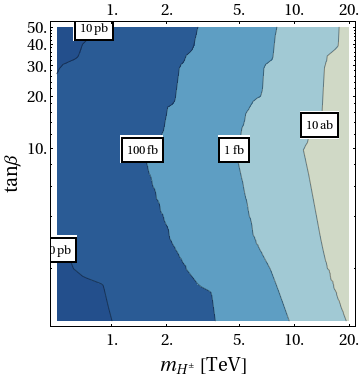}
\caption{$\sigma(pp \to tb H^\pm)$}
\label{fig:ProdHtb}
\end{subfigure}
\caption{Contours for the production cross-sections of neutral Higgs~\subref{fig:ProdHA}, neutral Higgs in association with two bottom quarks~\subref{fig:ProdHAbb} and charged Higgs in association with a bottom and a top quark~\subref{fig:ProdHtb} in the MSSM, at a \unit[100]{TeV} $pp$-collider.
}
\label{fig:xseccont}
\end{figure}

\begin{figure}
\begin{subfigure}{.32\textwidth}
\includegraphics[width=\textwidth]{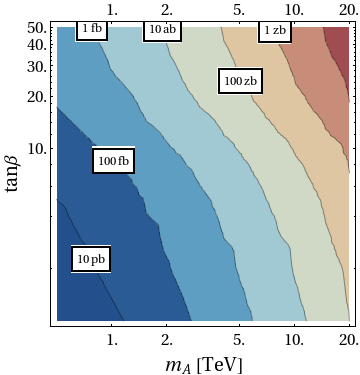}
\caption{$\sigma(pp \to H/A)\Br(H/A\to t \bar t)$}
\label{fig:XsecHA-tt}
\end{subfigure}
\hfill
\begin{subfigure}{.32\textwidth}
\includegraphics[width=\textwidth]{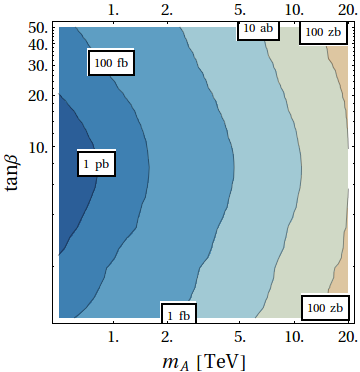}
\caption{$\sigma(pp \to b\bar bH/A)\Br(H/A\to t \bar t)$}
\label{fig:XsecHAbb-ttbb}
\end{subfigure}
\hfill
\begin{subfigure}{.32\textwidth}
\includegraphics[width=\textwidth]{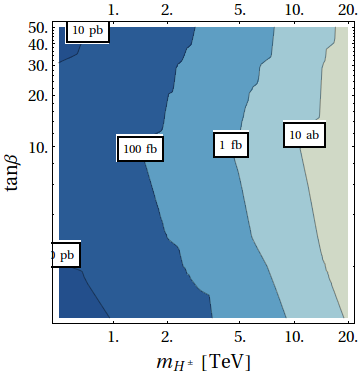}
\caption{$\sigma(pp \to t bH^\pm)\Br(H^\pm\to t b)$}
\label{fig:XsecHtb-ttbb}
\end{subfigure}
\caption{Contours for the cross-sections of neutral Higgs decaying to a top pair~\subref{fig:XsecHA-tt}, neutral Higgs produced in association with two bottom quarks and decaying to a top pair~\subref{fig:XsecHAbb-ttbb} and charged Higgs produced in association with a bottom and a top quark decaying to a bottom and a top quark~\subref{fig:XsecHtb-ttbb} in the MSSM, at a \unit[100]{TeV} $pp$-collider.
}
\label{fig:Xsall}
\end{figure}

\begin{figure}
\begin{subfigure}{.32\textwidth}
\includegraphics[width=\textwidth]{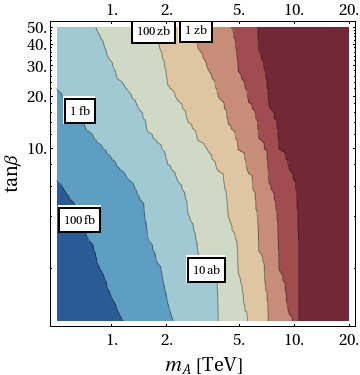}
\caption{$\sigma(pp \to H/A)\Br(H/A\to t \bar t)$}
\label{fig:XsecHA-tt-14}
\end{subfigure}
\hfill
\begin{subfigure}{.32\textwidth}
\includegraphics[width=\textwidth]{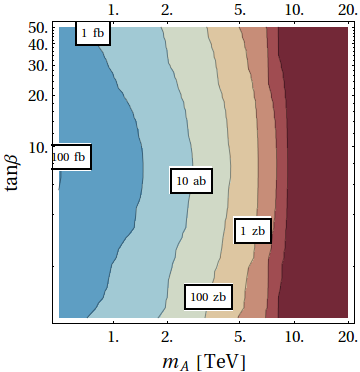}
\caption{$\sigma(pp \to b\bar bH/A)\Br(H/A\to t \bar t)$}
\label{fig:XsecHAbb-ttbb-14}
\end{subfigure}
\hfill
\begin{subfigure}{.32\textwidth}
\includegraphics[width=\textwidth]{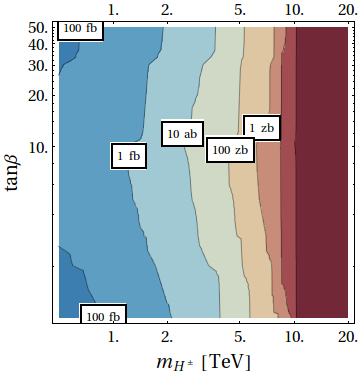}
\caption{$\sigma(pp \to t bH^\pm)\Br(H^\pm \to t b)$}
\label{fig:XsecHtb-ttbb-14}
\end{subfigure}
\caption{Contours for the cross-sections of neutral Higgs decaying to a top pair~\subref{fig:XsecHA-tt-14}, neutral Higgs produced in association with two bottom quarks and decaying to a top pair~\subref{fig:XsecHAbb-ttbb-14} and charged Higgs produced in association with a bottom and a top quark decaying to a bottom and atop quark~\subref{fig:XsecHtb-ttbb-14} in the MSSM, at the LHC.
}
\label{fig:Xsall-14}
\end{figure}

For similar reasons, the productions of these BSM Higgs bosons via vector boson fusion or Higgs Strahlung are highly suppressed.
Gluon fusion is the most important production mechanism of $H/A$ in the low $\tan\beta$ region, and $b b$ associated production becomes dominant in the moderate and large $\tan\beta$ regions (see Figure~\ref{fig:xseccont}).
The mild enhancement of $\sigma(pp \to H/A)$ in the large $\tan\beta$ region is mainly caused by the contribution from the $b$ quark loop.
Naturally we expect that $pp\to H/A \to t t$ and $pp\to b  b H/A \to bb \tau\tau,\ bbbb$ yield the best sensitivity in the low and large $\tan\beta$ regions, respectively.
One subtlety, however, arises from the interference between the signal and the QCD background.
It has been known for a while that the peak of the invariant mass $m_{tt}$ distribution for the $pp \to H/A \to tt$ can be distorted into a rather complicated peak–dip structure or even be  smoothed away, as $m_A$ increases~\cite{Frederix:2007gi, Dicus:1994bm}.
This potentially disables any resonance-reconstruction-based analysis, and necessitates the development of new strategies.
Such an exploration is beyond the scope of this paper. Instead, we propose to search in channels with $bb,tt$ associated Higgs productions, where the interference effects are less severe.

A remarkable observation however is: the moderate $\tan\beta$ region can be efficiently probed via the channel $pp \to b bH/A \to bbtt$, compared to the large and small $\tan\beta$ regions.
Explicitly, we present the contours for $\sigma(pp \to b bH/A)\Br(H/A\to t  t)$ at a \unit[100]{TeV} $pp$ collider in Figure~\ref{fig:Xsall}. For reference, we also present the contours of $\sigma(pp \to H/A)\Br(H/A\to t  t)$,  with the interference effects neglected.
Indeed, $\sigma(pp \to bbH/A)\Br(H/A\to t  t)$ is maximized in moderate $\tan\beta$ region.
Though its largest value for $m_A\sim \unit[1]{TeV}$ is one to two orders smaller than that of $\sigma(pp \to H/A)\Br(H/A\to t  t)$, they become comparable as $m_A$ increases towards $\sim \unit[10]{TeV}$.
Given that the channel $pp \to b b H/A \to bbtt$ carries more kinematic features due to the two additional $b$ quarks, we may well expect that this channel can yield a high sensitivity in probing for moderate $\tan\beta$ region.
This is true, as will be illustrated below.
A combination of these channels does leave no ``wedge'' open around moderate $\tan\beta$, given a $m_A$ within the energy reach of the LHC or the \unit[100]{TeV} $pp$ collider, if the low $\tan\beta$ region can be efficiently probed.

The $H^\pm$ production, it is dominated by $pp\to tb H^\pm$ in both large and small $\tan\beta$ regions (see Figure~\ref{fig:xseccont}).
This is simply because the coupling $g_{H^\pm tb}$ receives both $\tan\beta$-enhanced and $\tan\beta$-suppressed contributions at tree level.
Given that $H^\pm\to t b$ is the main decay mode in the decoupling limit, the channel $pp\to tb H^\pm\to tbtb$ is expected to be the  golden channel for the heavy $H^\pm$ searches.
The contours for $\sigma(pp \to bt H^\pm)\Br(H^\pm\to tb)$ at a \unit[100]{TeV} $pp$ collider is also presented in Figure~\ref{fig:Xsall}.
Because of the effects discussed above, if $m_A = m_{H^\pm}$, the largest $\sigma(pp \to bt H^\pm)\Br(H^\pm\to tb)$ is typically larger than the largest $\sigma(pp \to bbH/A)\Br(H/A\to t  t)$ and $\sigma(pp \to H/A)\Br(H/A\to t  t)$.
For example, the \unit[10]{ab} contour reaches $m_{H^\pm}\sim \unit[20]{TeV}$ for $H^\pm$, in comparison to $m_A\sim \unit[10]{TeV}$.

The contours for $\sigma(pp \to b b H/A)\Br(H/A\to t  t)$ and $\sigma(pp \to H/A)\Br(H/A\to t  t)$ and for $\sigma(pp \to bt H^\pm)\Br(H^\pm\to tb)$ in the MSSM at the LHC are presented in Figure~\ref{fig:Xsall-14}. As a comparison, the \unit[10]{ab} contours recede to $m_A\sim \unit[2]{TeV}$ and $m_{H^\pm} \sim \unit[4]{TeV}$, respectively. These contours can help understand the LHC and HL-LHC sensitivities in this regard which will be discussed below.

All cross-sections are calculated to leading order with \software[5.2.1.2]{MadGraph}~\cite{Alwall:2014hca}, in the case of signal cross-sections without applying any pre-cuts.
Additionally these results are crosschecked with \software[1.4.1]{SusHi}~\cite{Harlander:2012pb}.
Identical calculations for $\sqrt{s}=\unit[8]{TeV}$ match previous results for neutral~\cite{Spira:1995mt, Frank:2006yh, Djouadi:2013vqa} and charged Higgs~\cite{Djouadi:2013vqa, Flechl:2014wfa, Heinemeyer:2013tqa, Berger:2003sm}.
Our results for the branching ratios are calculated with \software{HDECAY}~\cite{Djouadi:1997yw}.
For our LHC analysis we apply a k-factor of 1.21 to the background~\cite{Cascioli:2013era}.

\section{Strategies for Collider Analyses}

\subsection{Collider Analyses and BDT}

In this article we mainly analyze the discovery reaches and exclusion limits of searches for both neutral and charged Higgs bosons at a \unit[100]{TeV} $pp$-collider. We cover the channels summarized in Table~\ref{tab:decay_tra}, except for the channel $pp \to bb H/A \to bbbb$, given that its coverage over parameter space largely overlaps with that of the channel $pp \to bb H/A \to bb\tau\tau$, and the channels $pp \to H/A \to tt$, $pp \to tt H/A \to tttt$, which are potentially sensitive in probing for the low $\tan\beta$ region. The corresponding backgrounds are summarized in Table~\ref{tab:XsecBackground}.
Because of their significance, we also present the $pp \to bb H/A \to bbtt$ and $pp \to tb H^\pm \to tbtb$ sensitivities that might be achieved at the LHC.

\begin{figure}
\begin{subfigure}{.32\textwidth}
\includegraphics[width=\textwidth]{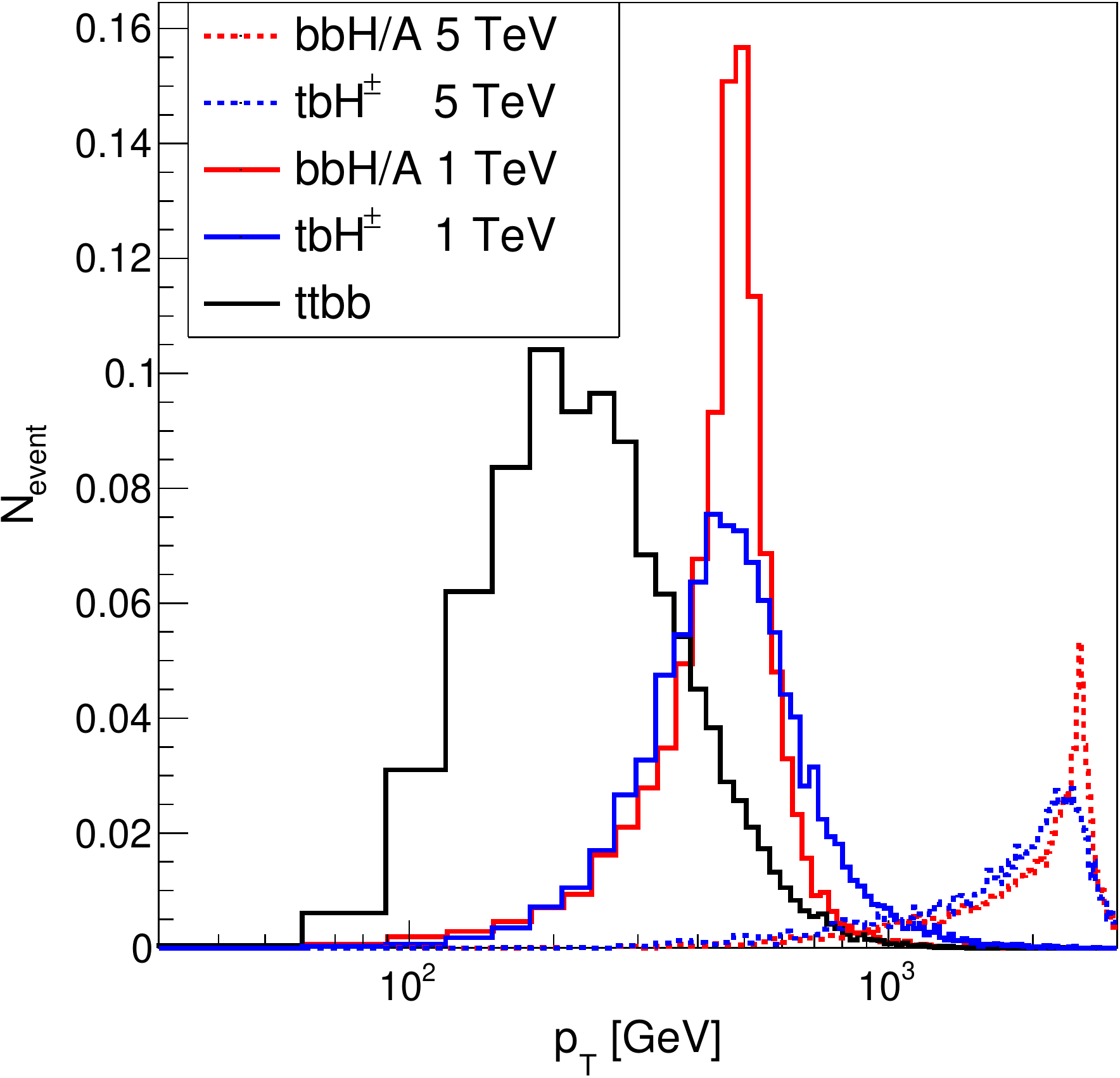}
\caption{Hardest $t$-quark}
\label{fig:toppt}
\end{subfigure}
\hfill
\begin{subfigure}{.32\textwidth}
\includegraphics[width=\textwidth]{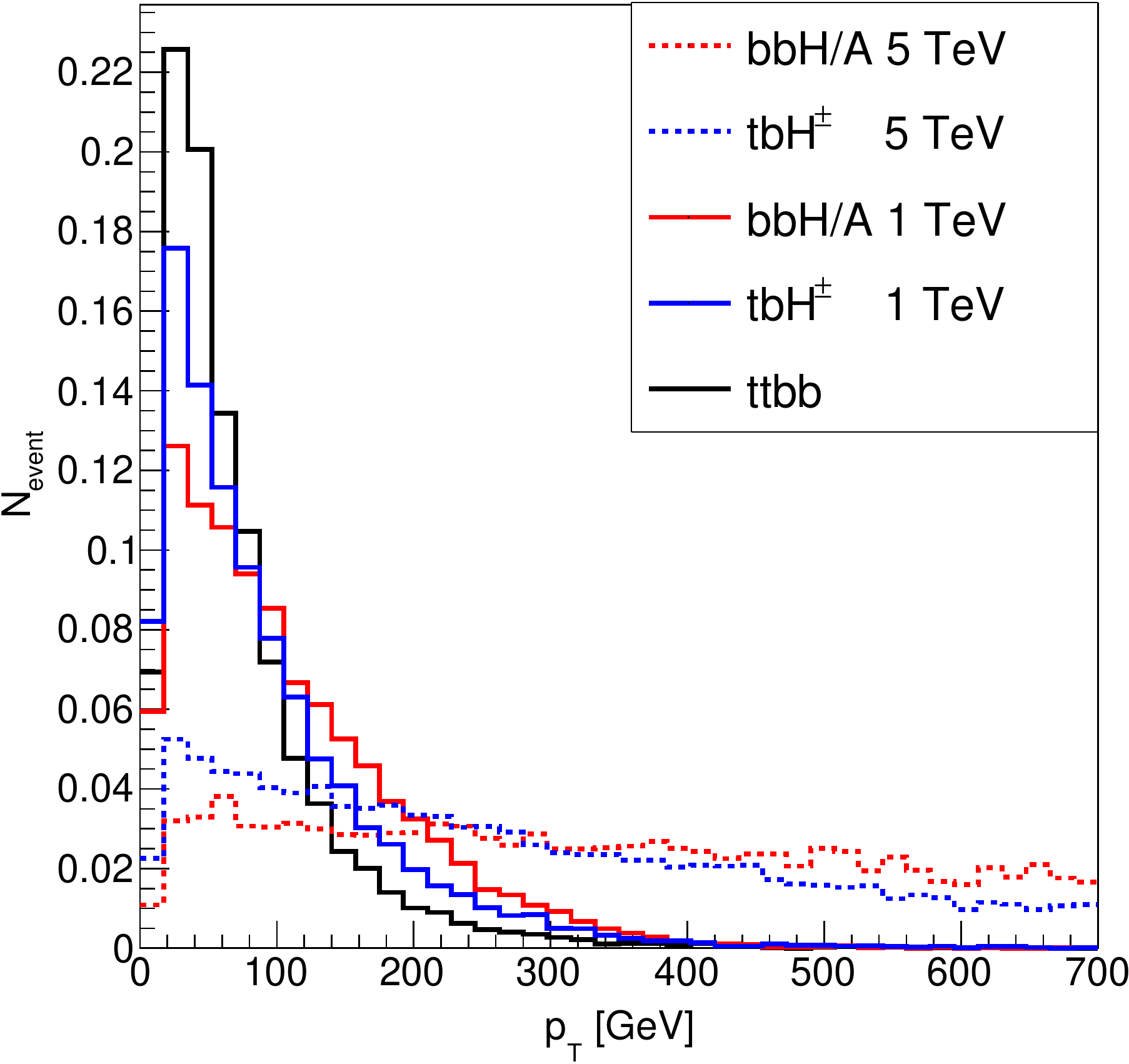}
\caption{Hardest lepton}
\label{fig:leptonpt}
\end{subfigure}
\hfill
\begin{subfigure}{.32\textwidth}
\includegraphics[width=\textwidth]{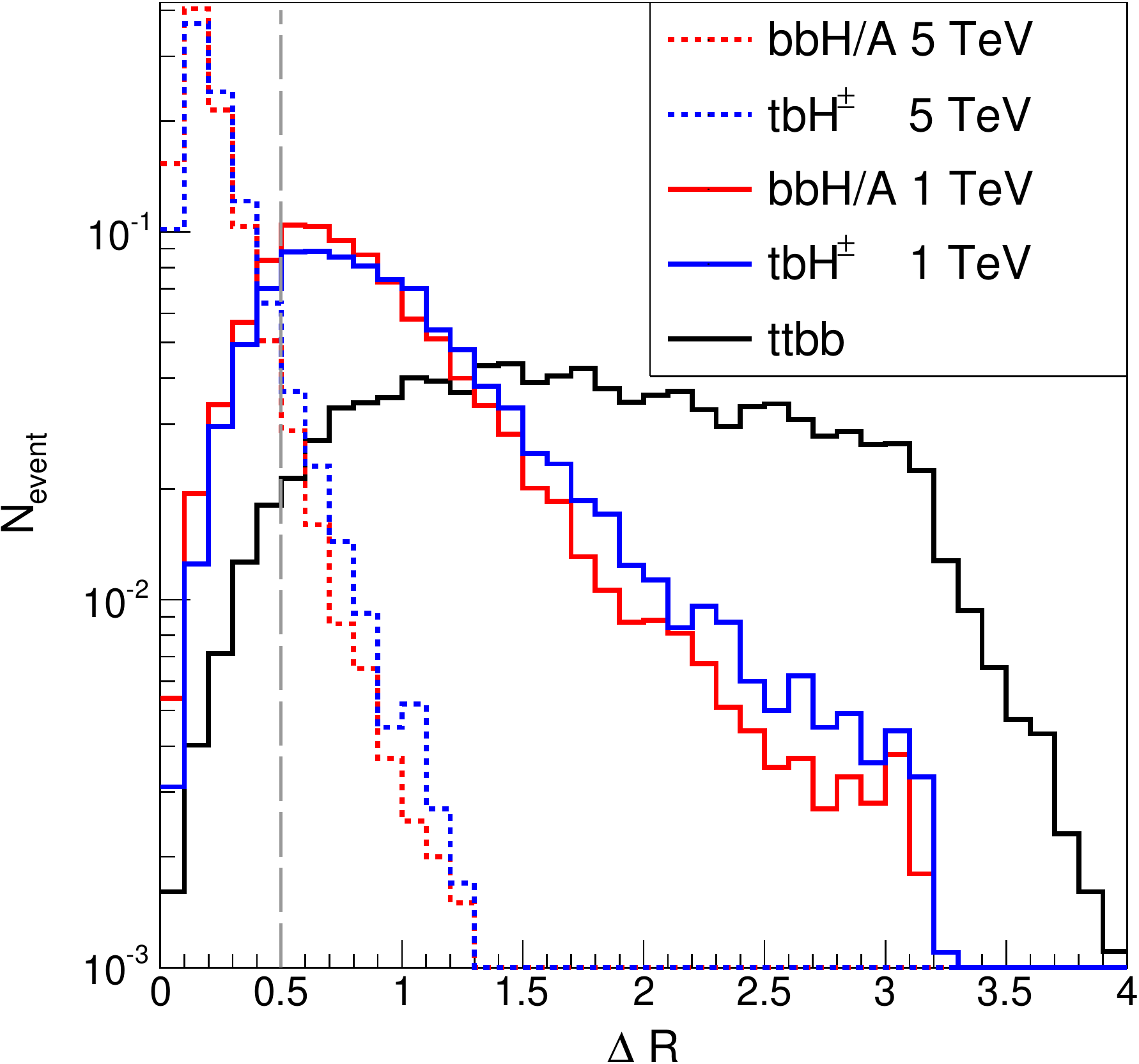}
\caption{Top opening angle}
\label{fig:TopDeltaR}
\end{subfigure}
\caption{%
Transverse momentum distribution of the hardest top~\subref{fig:toppt} and the hardest lepton~\subref{fig:leptonpt} for background ($ttbb$) and signal ($bbH/A$, $tbH^\pm$).
Additionally, we show the opening angle between lepton and $b$-quark coming from the decay of the hardest top~\subref{fig:TopDeltaR}.
The vertical line at $\Delta R = 0.5$ represents the jet radii we have chosen.
The areas under all curves are normalized.}
\label{fig:HardestPt}
\end{figure}

\begin{figure}
\hfill
\begin{subfigure}[t]{.32\textwidth}
\includegraphics[width=\textwidth]{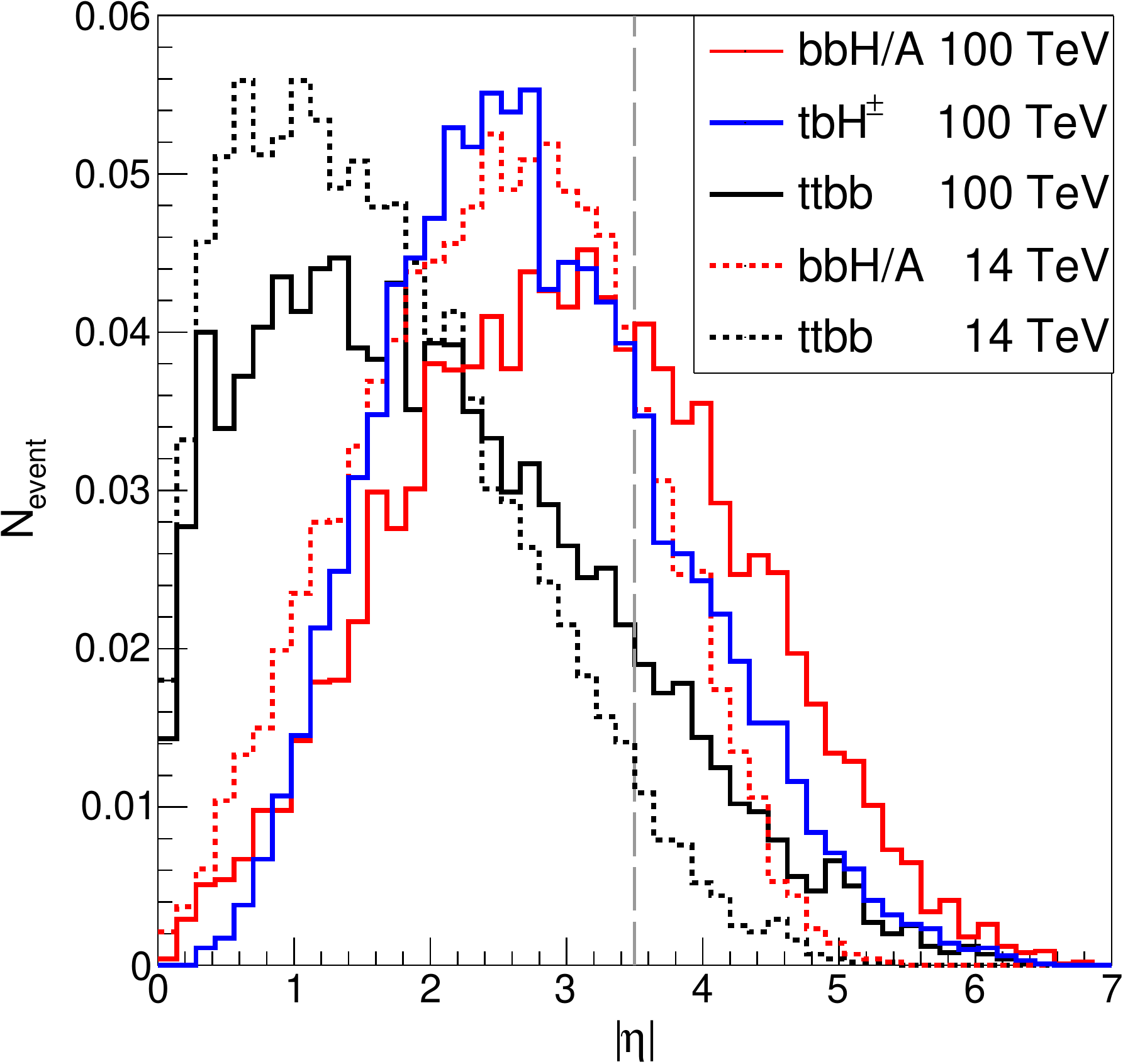}
\caption{$b$-quarks accompanying Higgs production}
\label{fig:pseudorapidity}
\end{subfigure}
\hfill
\begin{subfigure}[t]{.32\textwidth}
\includegraphics[width=\textwidth]{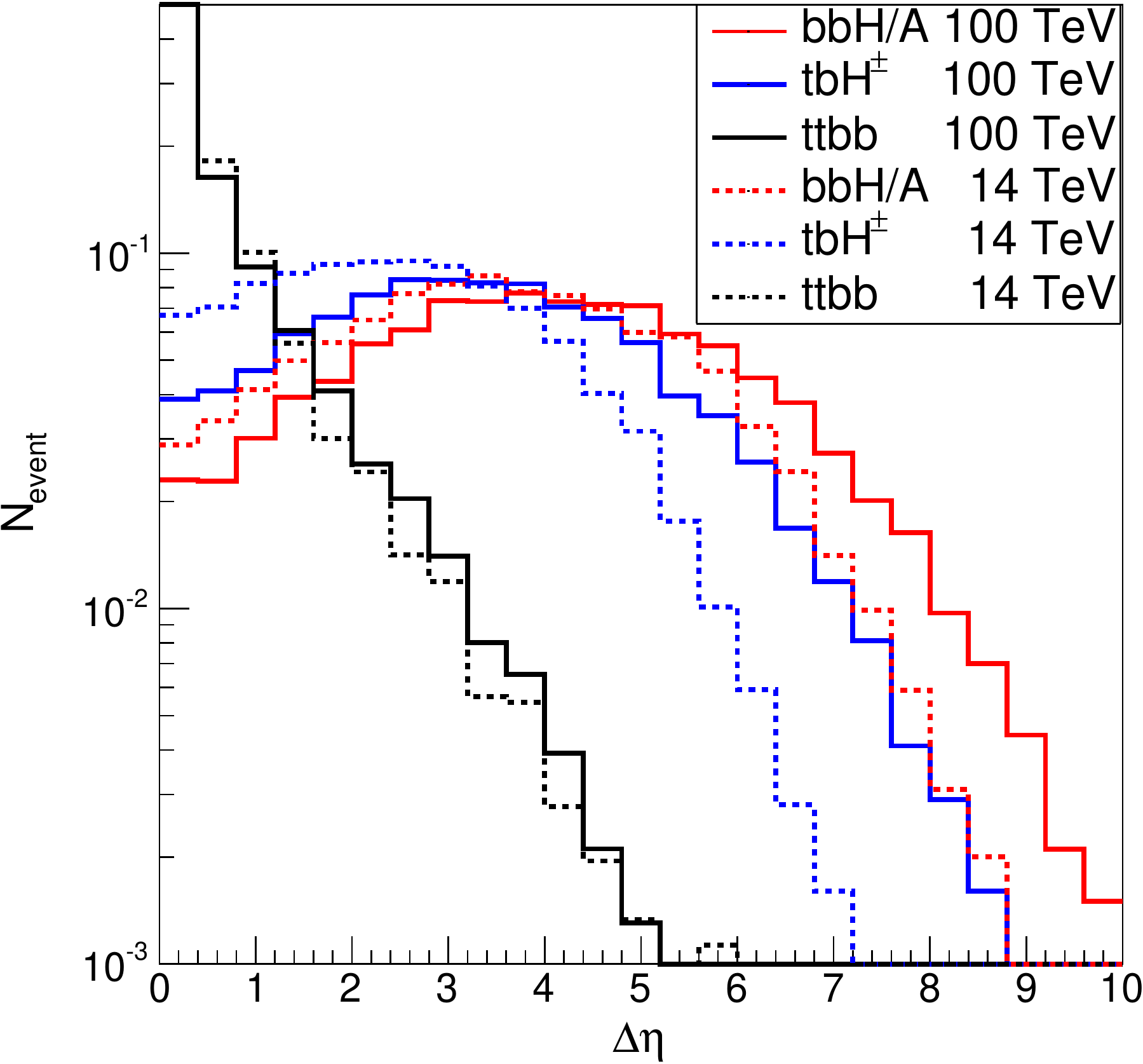}
\caption{$\Delta\eta$ between the two accompanying $b$-quarks}
\label{fig:pseudorapiditydifference}
\end{subfigure}
\hfill\strut
\caption{Rapidity distributions of the $b$-quarks accompanying Higgs production~\subref{fig:pseudorapidity} and the distribution of the difference in rapidity between these two $b$-quarks~\subref{fig:pseudorapiditydifference}.
The graphs show background ($ttbb$) and signal ($bbH/A$, $tbH^\pm$) for \unit[14]{TeV} and \unit[100]{TeV}. We choose the $b$-quarks accompanying the top production in $ttbb$ for comparison. For \unit[14]{TeV} and \unit[100]{TeV} a cut on the jet transverse momentum of \unit[20]{GeV} and \unit[40]{GeV} has been applied, respectively.
The vertical line at $|\eta| = 3.5$ in~\subref{fig:pseudorapidity} represents the tracker coverage assumed for an \unit[100]{TeV} $pp$-collider, for jets with larger rapidity $b$-tagging is not applicable. The areas under all curves are normalized.
}
\label{fig:EtaDistribution}
\end{figure}

The analyses are largely based on two classes of kinematic features. The first one is related to the heaviness of Higgs bosons in the decoupling limit.
In such a case, the decay products of Higgs are strongly boosted, resulting e.g. for $H/A \to tt$ in two boosted tops and
for $H^\pm\to tb$ in one boosted top. In the analyses with intermediate top quarks, we assume that the more boosted one decays leptonically
and optimize the analyses based on this assumption.
The distribution of the hardest top-quark and the hardest lepton are depicted in Figure~\ref{fig:toppt} and~\ref{fig:leptonpt}, respectively.
The decay products of these tops tend to lie within a single jet cone (cf.~Figure~\ref{fig:TopDeltaR}), yielding a top jet.
Additionally, the events have a large sum of scalar transverse momenta ($H_T$). These features becomes less prominent for smaller Higgs masses, especially for the mass range accessible to the LHC.

Another class of kinematic features is related to the particles accompanying the Higgs production. For example, in the case of $bbH/A$, the two $b$-jets accompanying the Higgs are less boosted, but tend to have a large rapidity.
Hence, the difference in rapidity ($\Delta \eta$) is typically larger than that of $b$-jet pairs in the main backgrounds $ttbb$ and $ttcc$.
Therefore, $\Delta \eta$ can serve as a strong discriminator if the two accompanying $b$-jets are identified. The story is similar in the case of $tb H^\pm$, where the second b-jet is from top decay. However, the rapidity of many of these $b$-jets is larger than the coverage of the tracker of a LHC-like detector design ($|\eta| < 2.5$), as is indicated in Figure~\ref{fig:EtaDistribution}.
Hence it is difficult for these $b$-jets to be identified. Even if these $b$-jets are tagged, we are still confronted by a combinatorial background caused by $b$-jets from Higgs decays which are typically central and hence easier to be tagged.

\begin{figure}
\centering
\begin{subfigure}{.9\textwidth}
\centering
\tikzsetnextfilename{FlowChartBoosted}
\footnotesize
\input{figs/FlowChartBoosted.pgf}
\caption{BDT-based event reconstruction in the case of boosted tops.}
\label{fig:flow chart boosted}
\end{subfigure}
\par\bigskip
\centering
\begin{subfigure}{.9\textwidth}
\centering
\tikzsetnextfilename{FlowChart}
\footnotesize
\input{figs/FlowChart.pgf}
\caption{BDT-based event reconstruction in the case of  non-boosted tops.}
\label{fig:flow chart unboosted}
\end{subfigure}
\caption{%
Steps undertaken during the BDT-based reconstruction of the signal $pp \to H/A bb \to ttbb$.
Based on the EFlow output from \software{Delphes} we cluster jets with \software{FastJet}.
Solid lines depict a reconstruction step where we combine more than one objects to form a new object.
Dashed lines symbolize tagging steps.
}
\label{fig:flow charts}
\end{figure}
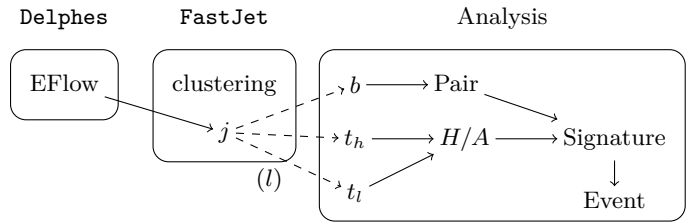
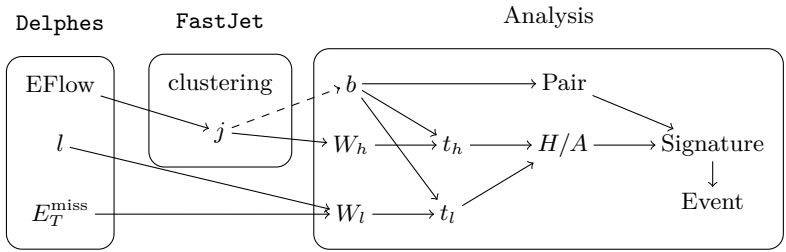

Addressing these difficulties is not easy, however, it provides an opportunity to explore potential guidelines for optimizing the detector design of an \unit[100]{TeV} $pp$-collider, as well as for searching for heavy resonances at such a collider, given the essential role played by an extended Higgs sector in particle physics.
In this article, we assume a tracker coverage of $|\eta| < 3.5$ for future \unit[100]{TeV} collider detectors.
In response to the boostness kinematics and in order to suppress combinatorial background, we apply a multivariate approach by using a BDT.
More explicitly, we apply BDT in each reconstruction step, aiming for a full event reconstruction. Below are the steps:
\begin{enumerate}
\item Construct bottom BDT. With a BDT method, we are able to define $b$-like jets (similar for $t$-like jets) which are characterized by their likelihood to be a $b$-jet.

\item Construct top BDTs: one is hadronic and another one is leptonic. We build up BDT-based jet-taggers for boosted tops which mostly originate from the Higgs decays (cf.~Figure~\ref{fig:TopDeltaR}).
In the case of less boosted tops we can not expect all top decay products to lie inside a jet cone and we have to reconstruct the $W$-bosons before we can use these to reconstruct the tops.

\item Construct the Higgs BDT using the reconstructed tops or one reconstructed top and one $b$-like jet for neutral or charged Higgs, respectively.

\item Construct a BDT to demand two $b$-like jets (or one top-like and one bottom-like jet) with large $\Delta \eta$ which accompany the Higgs production. We name it ``bottom-fusion'' BDT, though such kinematics may originate from a process other than bottom quark fusion.

\item Construct the BDT for the whole event by properly combining the Higgs BDT and the bottom-fusion BDT.
\end{enumerate}
For illustration, the reconstruction steps undertaken for the $pp \to bb H/A \to bbtt$ events with heavily boosted top-pair and unboosted top-pair are depicted in Figure~\ref{fig:flow chart boosted} and in Figure~\ref{fig:flow chart unboosted}, respectively.
The transition from boosted to unboosted events is gradual, which is taken into consideration by the top BDT.

In the large $\tan\beta$ region, the decays into $\tau$-leptons contribute to the searches for both neutral and charged Higgs bosons.
We require the two $\tau$ leptons in $pp \to bb H/A \to bb\tau\tau$ to decay semi-leptonically. In $pp \to tb H^\pm \to tb\tau\nu$, we let both top quark and $\tau$ decay hadronically.
Then we can use hard leptons from the $\tau$ decay or large transverse mass to suppress backgrounds in these two cases respectively.
Given their relatively simple collider kinematics, we apply a cut-based method in these two cases.

Our analysis framework is defined in the following. We use \software[3.5.1]{SoftSusy}~\cite{Allanach:2001kg} as generator for the MSSM Higgs spectrum and simulate all events with \software{MadGraph}.
The subsequent decays are performed by \software[6.426]{Pythia}~\cite{Sjostrand:2006za}.
We use \software[3.1.2]{Delphes}~\cite{deFavereau:2013fsa} to simulate a detector with CMS geometry (except using a tracker coverage $|\eta| < 3.5$ for the \unit[100]{TeV} $pp$-collider), with pile-up turned off.
For LHC simulations we keep the default values.
For cut-based analyses we depend on \software{Delphes} jet-tagging, which we tune according to Drell-Yan samples with a transverse momentum cut of \unit[40]{GeV}.
The $b$-tagging efficiency is \unit[70]{\%} with \unit[20]{\%} misidentification of $c$-jets.
The $\tau$-tagging efficiency is \unit[60]{\%} and the fake-rate is \unit[1]{\%}.
For BDT-based analyses we cluster jets based  on \software{Delphes} energy flow observables with \software[3.0.6]{FastJet}~\cite{Cacciari:2011ma}, with a radius 0.5.
In this case we switch off lepton isolation requirements, and instead require that the leading lepton has a transverse momentum larger than 50 GeV and 100 GeV, and that the missing energy is larger than 30 GeV and 60 GeV, for the LHC and the \unit[100]{TeV} collider, respectively.
We demand that jets have a transverse momentum larger than \unit[20]{GeV} and \unit[40]{GeV} for the LHC and the \unit[100]{TeV} collider, respectively, in all cases.
 In order to reduce the number of background events, we apply a pre-cut on the top or lepton transverse momenta or the missing transverse energy (cf. Table~\ref{tab:XsecBackground}).
The background cross-sections after all pre-cuts are summarized in Table~\ref{tab:XsecBackground}.
We use the \software[4.2.0]{TMVA} package~\cite{Hocker:2007ht} of the \software[5.34]{ROOT} framework~\cite{Brun:1997pa} for BDT analyses.

\subsection{BDT-based Top-jet Taggers}

\begin{figure}
\begin{subfigure}{.49\textwidth}
\includegraphics[width=\textwidth]{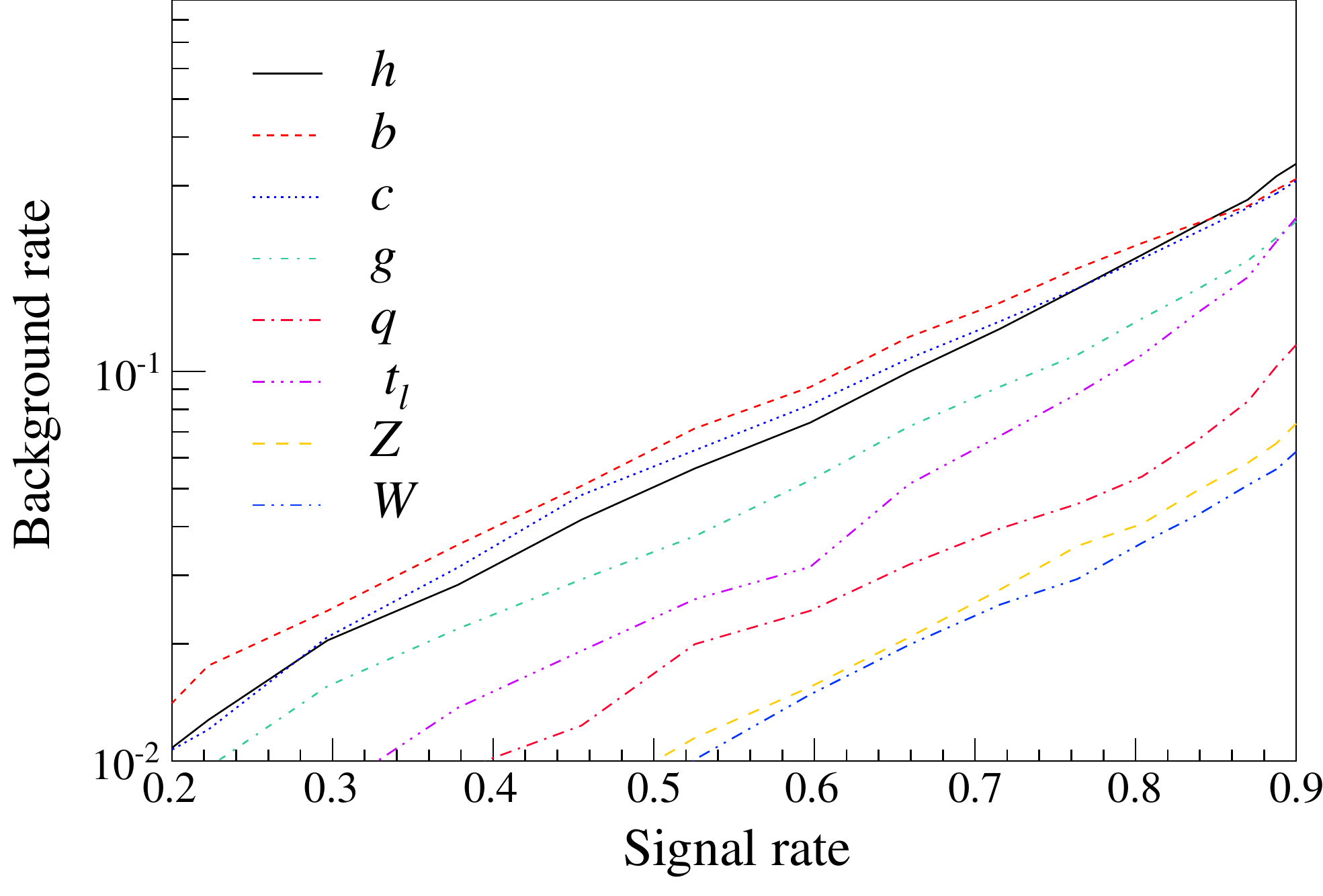}
\caption{$\unit[700]{GeV} < p_T^j(\text{hadronic}) < \unit[1000]{GeV}$}
\label{fig:hadronic top tagger 700}
\end{subfigure}
\hfill
\begin{subfigure}{.49\textwidth}
\includegraphics[width=\textwidth]{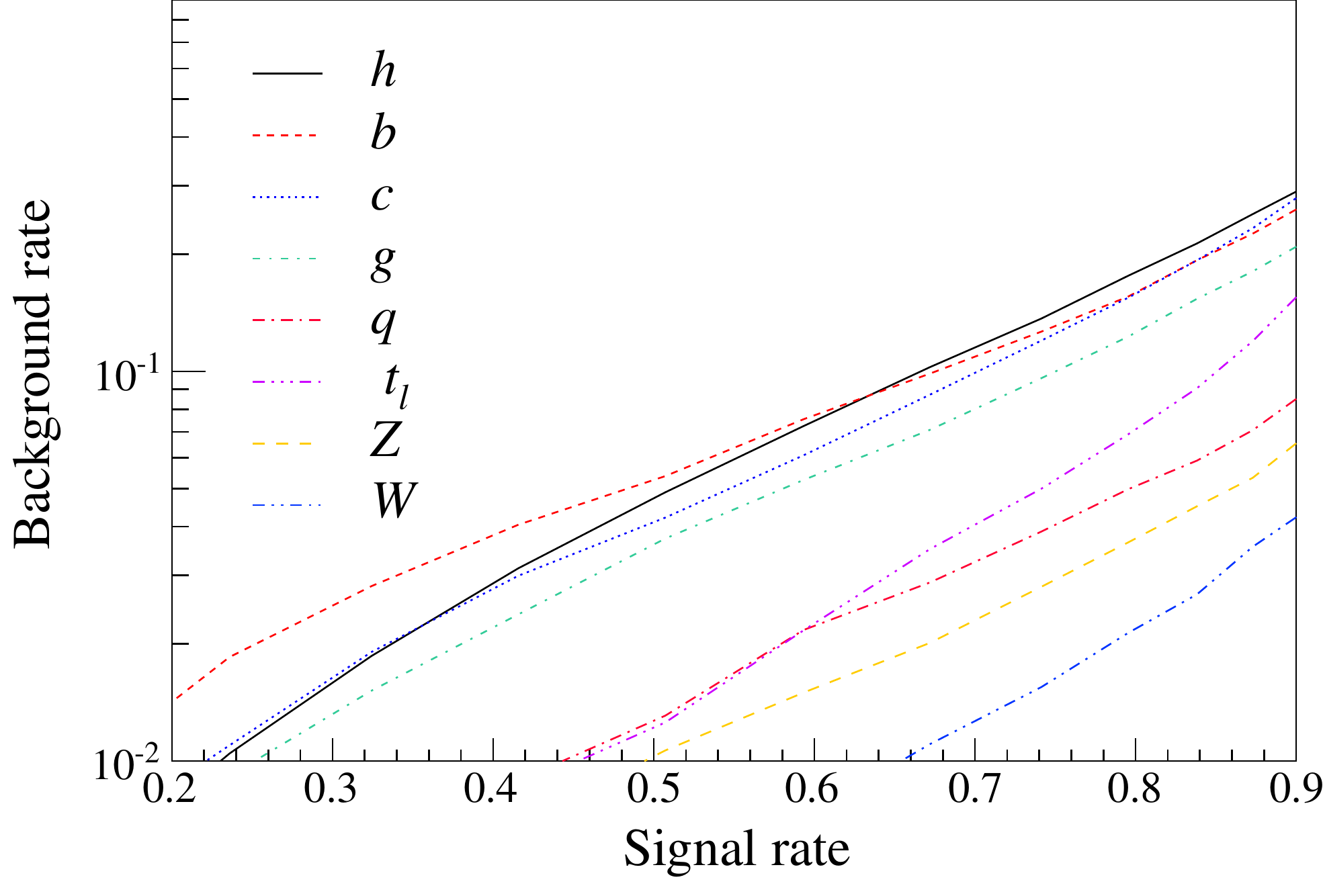}
\caption{$\unit[1000]{GeV} < p_T^j(\text{hadronic}) < \unit[1500]{GeV}$}
\label{fig:hadronic top tagger 1000}
\end{subfigure}
\par\bigskip
\begin{subfigure}{.49\textwidth}
\includegraphics[width=\textwidth]{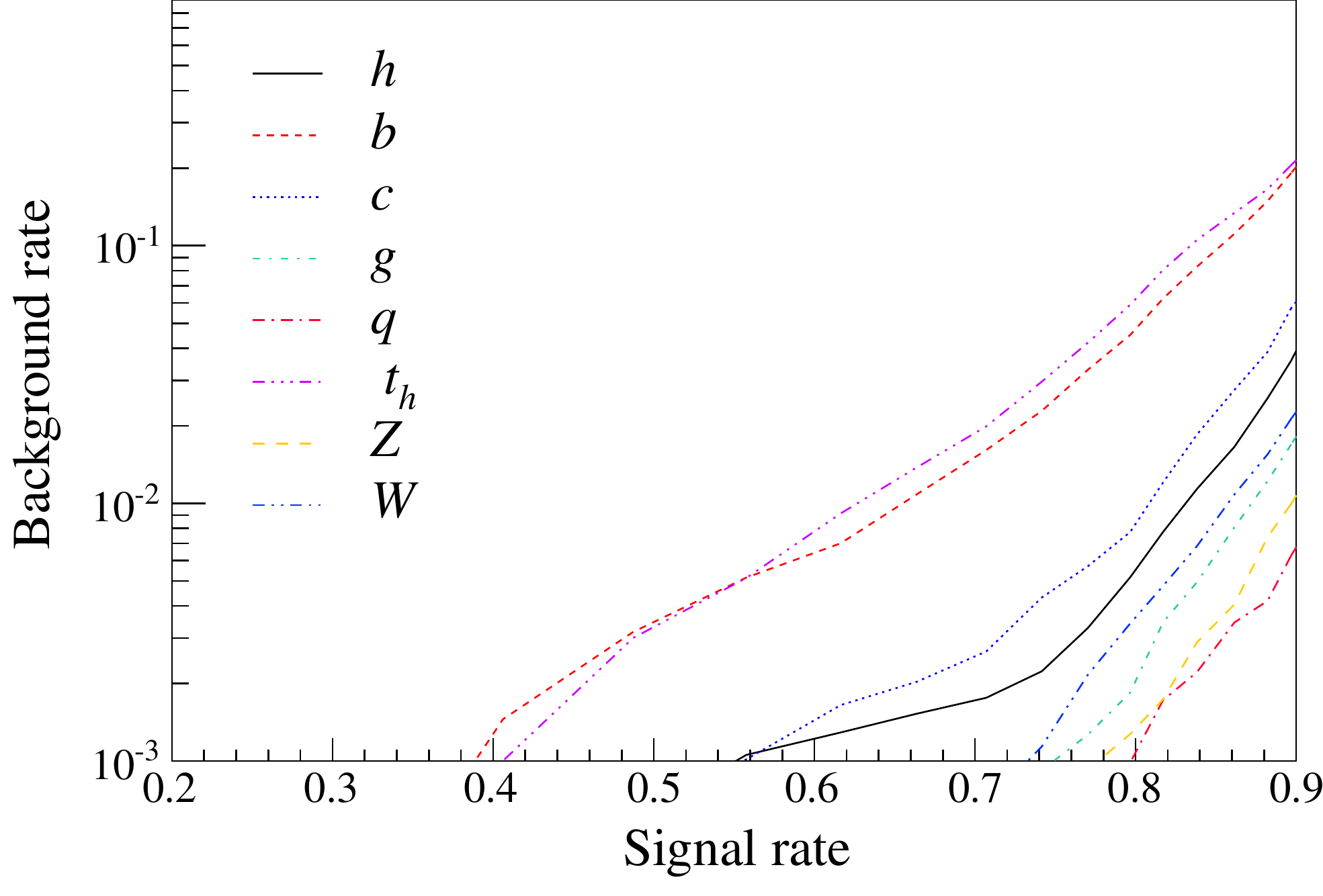}
\caption{$\unit[500]{GeV} < p_T^j(\text{leptonic}) < \unit[1000]{GeV}$}
\label{fig:leptonic top tagger 500}
\end{subfigure}
\hfill
\begin{subfigure}{.49\textwidth}
\includegraphics[width=\textwidth]{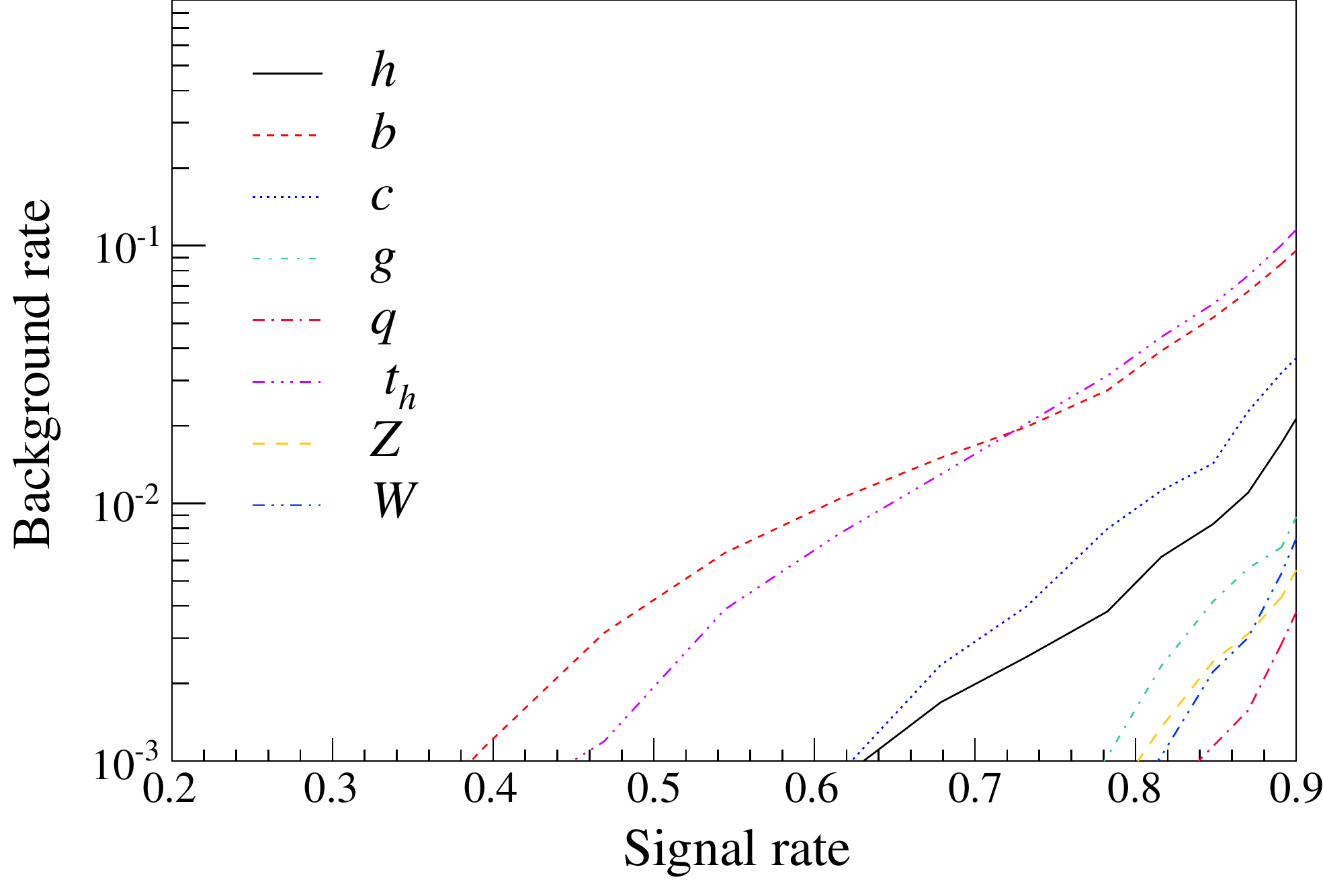}
\caption{$\unit[1000]{GeV} < p_T^j(\text{leptonic}) < \unit[1500]{GeV}$}
\label{fig:leptonic top tagger 1000}
\end{subfigure}
\caption{Misidentification rate of the hadronic~\subref{fig:hadronic top tagger 700} and~\subref{fig:hadronic top tagger 1000} as well as leptonic~\subref{fig:leptonic top tagger 500} and~\subref{fig:leptonic top tagger 1000} top-taggers as functions of the tagging rate.
We consider only the case of highly boosted top jets and do not use sub-jet information.
Quark-jets (including $t$, $b$, $c$ and $q=u,d$) are produced in a Drell-Yan process; $g$-jets are produced via QCD processes; the SM $h$-jets are produced via $gg\to hh$, with Br$(h\to bb)=100\%$ assumed; and $W$- and $Z$-jets are produced via di-boson production, and decay in a standard way.  All samples are subject to a pre-cut on their transverse momenta, and are generated in the same size.
We present two choices of pre-cuts for each tagger. In the hadronic case we generate the jets with a parton level pre-cut of \unit[500]{GeV} and \unit[1000]{GeV} and additionally require the jets to fall into the transverse momentum windows $\unit[700]{GeV} < p_T^j < \unit[1000]{GeV}$ and $\unit[1000]{GeV} < p_T^j < \unit[1500]{GeV}$, respectively.  In the leptonic case we generate the jets with a parton level pre-cut of \unit[500]{GeV} and \unit[1000]{GeV} and additionally require the jets to fall into the transverse momentum windows $\unit[500]{GeV} < p_T^j < \unit[1000]{GeV}$ and $\unit[1000]{GeV} < p_T^j < \unit[1500]{GeV}$, respectively.
We do not consider the effects caused by pile-up.
}
\label{fig:hadronic top tagger}
\end{figure}

Each top BDT comprises a boosted top-jet tagger and modules for the reconstruction of unboosted tops.
We apply both techniques to each top and rely on the technique with the better result, hence we choose the top jet with the best BDT response.
The construction of top-jet taggers plays a crucial role in exploring physics with boosted tops.
Thus we introduce in this section how to construct the hadronic and leptonic top-jet taggers, using a BDT method.%
\footnote{During the preparation of this article a BDT improved version of the HEPTopTagger has been published~\cite{Kasieczka:2015jma}.}

For heavily boosted hadronic top jets the detector resolution may not allow for using sub-jet information. Hence we rely solely on the jet mass
and variables making use of the secondary vertex information (cf. Appendix~\ref{sec:boosted decision trees}).
In order to suppress the fake rate of leptonically decaying tops in the hadronic top tagger,
we veto against a hard lepton inside the jet cone. If the top-jets are not too heavily boosted, we make use of the kinematics of sub-jets~\cite{Plehn:2009rk}, which are re-clustered using a $k_T$ jet algorithm in an exclusive way.
We consider the case, where the top-jet consists of two sub-jets which resemble a boosted $W$-jet and a $b$-jet respectively, and the case, where the top-jet consists of three sub-jets%
\footnote{To build up the full top BDT, we incorporate two additional cases, where the reconstructed top consists of three separated jets, and where the reconstructed top consists of two separated jets with one resembling a boosted $W$-jet and another one resembling a $b$-jet.}.
We assume that the cell resolution of a typical detector is around $\Delta R \simeq 0.1$ which imposes an upper limit on the substructure resolution.
We rely solely on traditional kinematic variables of the jets and sub-jets.
This can be further improved with variables based on more advanced substructure information, such as pull~\cite{Gallicchio:2010sw} or di-polarity~\cite{Hook:2011cq}, which should help to suppress the fake rate of jets with a different color-flow compared to top jets.

In the case of leptonic top jets, it becomes crucial to identify the lepton together with the hadronic activity.
As the lepton is not isolated it has to serve as the foundation for a boosted leptonic top tagger~\cite{Cohen:2014hxa}, which poses more challenges in the case of electrons compared to muons.
The development of the necessary non-isolated lepton taggers lies out of the scope of this paper.
Therefore, we have assumed that it will be possible to identify leptons in a hadronic environment.

We have tested these top taggers on various backgrounds. The resulting misidentification rates as a function of the top-tagging rate are presented in Figure~\ref{fig:hadronic top tagger}. We present the miss-identification rate for two bins each in which all jets are generated with the same parton level pre-cut of $p_T^j > \unit[500 \text{ and } 1000]{GeV}$, respectively.
Additionally, we require the jets to fall into a transverse momentum window of $\unit[500, 700]{GeV} < p_T^j < \unit[1000]{GeV}$ and $\unit[1000]{GeV} < p_T^j < \unit[1500]{GeV}$, respectively.
We apply a slightly different lower cut on the lower of the two $p_T$ windows for leptonic (\unit[500]{GeV}) and hadronic (\unit[700]{GeV}) tagger, in order to compensate for the missing transverse momentum in the leptonic case. For the hadronic top tagger the most important background consists of Higgs jets followed by $b$, $c$ and $h$ jets. For the leptonic top tagger the most important backgrounds are hadronic top and $b$ jets.
We stress that the faking rates shown in Figure~\ref{fig:hadronic top tagger} are defined inclusively, hence the BDT background is defined to a combination of all listed background jets.
At an exclusive level, the fake rates can be further optimized.

\section{\texorpdfstring{Prospects at \unit[14 and 100]{TeV} \boldmath$pp$ Colliders}{Prospects at 14 and 100 TeV pp Colliders}}
\label{sec:ana}

\begin{figure}
\begin{subfigure}{.49\textwidth}
\includegraphics[width=\textwidth]{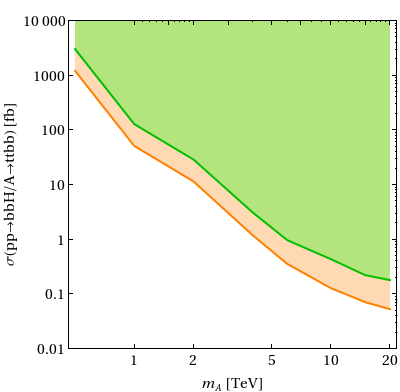}
\caption{Neutral Higgs at \unit[100]{TeV}}
\label{fig:ModelIndependtNeutral1}
\end{subfigure}
\hfill
\begin{subfigure}{.49\textwidth}
\includegraphics[width=\textwidth]{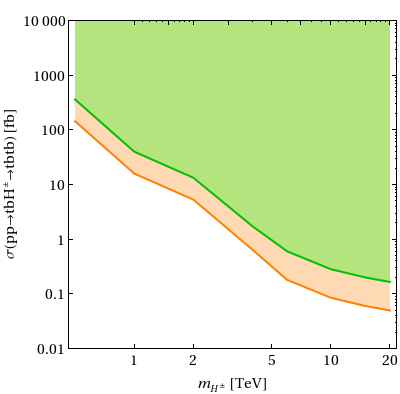}
\caption{Charged Higgs at \unit[100]{TeV}}
\label{fig:ModelIndependtCharged1}
\end{subfigure}
\begin{subfigure}{.49\textwidth}
\includegraphics[width=\textwidth]{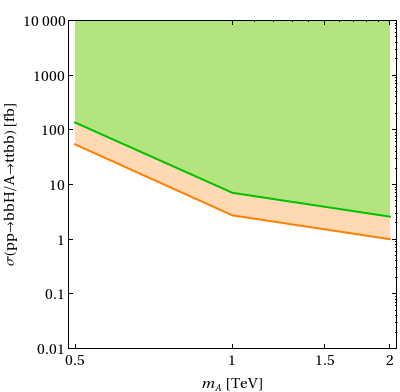}
\caption{Neutral Higgs at \unit[14]{TeV}}
\label{fig:ModelIndependtNeutral2}
\end{subfigure}
\hfill
\begin{subfigure}{.49\textwidth}
\includegraphics[width=\textwidth]{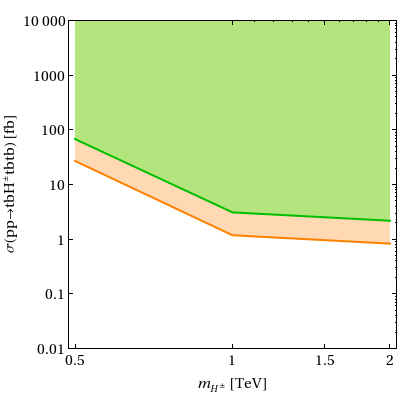}
\caption{Charged Higgs at \unit[14]{TeV}}
\label{fig:ModelIndependtCharged2}
\end{subfigure}
\caption{Model independent reaches of $bbH/A \to t_ht_lbb$ at~\subref{fig:ModelIndependtNeutral1} and~\subref{fig:ModelIndependtNeutral2} as  well as $tb H^\pm \to t_h b t_l b$ in~\subref{fig:ModelIndependtCharged1} and~\subref{fig:ModelIndependtCharged2}.
In the upper figures~\subref{fig:ModelIndependtNeutral1} and~\subref{fig:ModelIndependtCharged1} the energy is \unit[100]{TeV}.
In the lower figures~\subref{fig:ModelIndependtNeutral2} and~\subref{fig:ModelIndependtCharged2} the energy is \unit[14]{TeV}.
The Luminosity in all figures is $\unit[3]{ab^{-1}}$.
The green area represents the discovery reach and the orange band shows the exclusion reach.
The constraints are weaker at the \unit[100]{TeV} collider compared to the LHC, because of the bigger background cross-sections.
}
\label{fig:modelind}
\end{figure}

\begin{figure}
\begin{subfigure}{.49\textwidth}
\includegraphics[width=\textwidth]{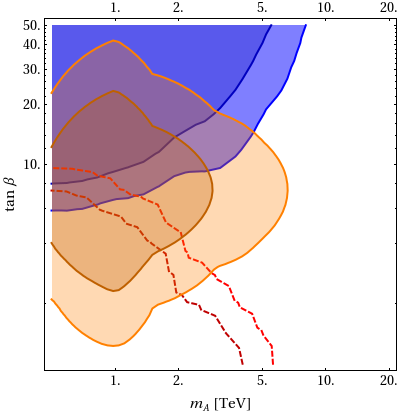}
\caption{Discovery reach for neutral Higgs}
\label{fig:discoveryneutral}
\end{subfigure}
\hfill
\begin{subfigure}{.49\textwidth}
\includegraphics[width=\textwidth]{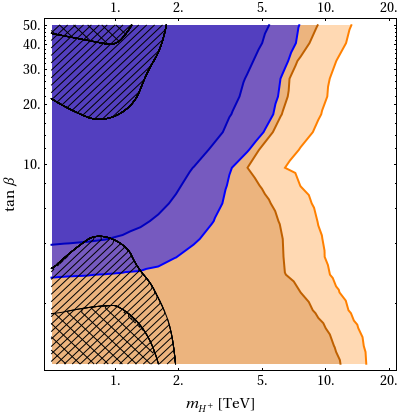}
\caption{Discovery reach for charged Higgs}
\label{fig:discoverycharged}
\end{subfigure}
\begin{subfigure}{.49\textwidth}
\includegraphics[width=\textwidth]{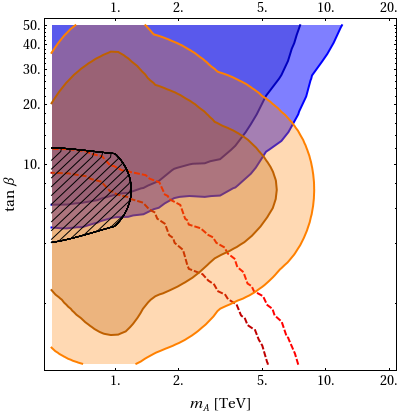}
\caption{Exclusion reach for neutral Higgs}
\label{fig:exclusionneutral}
\end{subfigure}
\hfill
\begin{subfigure}{.49\textwidth}
\includegraphics[width=\textwidth]{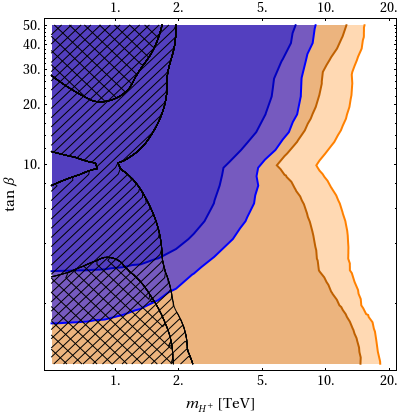}
\caption{Exclusion reach for charged Higgs}
\label{fig:exclusioncharged}
\end{subfigure}
\caption{Discovery reaches~\subref{fig:discoveryneutral} and~\subref{fig:discoverycharged} and exclusion limits~\subref{fig:exclusionneutral} and~\subref{fig:exclusioncharged} of the MSSM Higgs bosons.
For the neutral Higgs bosons~\subref{fig:discoveryneutral} and~\subref{fig:exclusionneutral}, the blue and  orange regions are probed for by the channels $pp \to bb H/A \to bb \tau_h\tau_l$, $pp \to bb H/A \to bb t_h t_l$, respectively.
For the charged Higgs bosons~\subref{fig:discoverycharged} and~\subref{fig:exclusioncharged}, the blue and orange regions are probed by the channels $pp \to tb H^\pm \to tb \tau_h\nu$ and $pp \to tb H^\pm \to t_h b t_l b$, respectively.
Given the same channel or the same color, the two different opacities indicate the sensitivities w.r.t. a luminosity of $\unit[3]{ab^{-1}}$ and $\unit[30]{ab^{-1}}$ at a \unit[100]{TeV} $pp$ collider, respectively.
The cross-hatched and diagonally hatched regions are the predicted discovery contours (or exclusion contours) for associated Higgs production at the LHC for $\unit[0.3]{ab^{-1}}$, and $\unit[3]{ab^{-1}}$, respectively.
}
\label{fig:HA}
\end{figure}

The exclusion limits and discovery reaches which are yielded by these channels at the LHC and a \unit[100]{TeV} $pp$ collider with a CMS-like detector (a tracker coverage of $|\eta|<3.5$ is assumed for the 100 TeV machine) are presented in Figure~\ref{fig:modelind} and Figure~\ref{fig:HA}.
We provide limits for the LHC with an integrated Luminosity of $\unit[0.3]{ab^{-1}}$ and $\unit[3]{ab^{-1}}$ which corresponds to the expectation of the collected data after the third run and an upgrade to the HL-LHC, respectively.
For the \unit[100]{TeV} collider we show the contours for a luminosity of $\unit[3]{ab^{-1}}$ and $\unit[30]{ab^{-1}}$.
Though systematic errors are not considered, we believe that incorporating them will not qualitatively change conclusions reached in this paper.

For one to interpret these results straightforwardly in some other extended Higgs sectors, e.g., the one in the 2HDM, we present the exclusion limits and discovery reaches of the channels $bbH/A \to t_ht_l$ and $tb H^\pm \to t_h b t_l b$ at \unit[14]{TeV} and \unit[100]{TeV} in Figure~\ref{fig:modelind}, in a model-independent way.
A generic feature of these sensitivity reaches is that they tend to be higher, for a larger $m_A$ or $m_{H^\pm}$ within the range under exploration.
This is simply because the boostness kinematics plays a crucial role in suppressing the background.

The interpretation of these results in the MSSM are presented in Figure~\ref{fig:HA}.
For moderate $\tan\beta$, the cross-section for the associated charged Higgs channel (cf.~Figure~\ref{fig:XsecHtb-ttbb}) and the associated neutral Higgs channel (cf.~Figure~\ref{fig:XsecHAbb-ttbb}) are comparable to each other.
Therefore, the analyses reach a roughly comparable sensitivity for this parameter point, given that they share similar backgrounds for the setup under consideration.
In the case of charged Higgs the cross section and the sensitivity increases when $\tan\beta$ moves away from this point.
In the case of neutral Higgs the cross-section and sensitivity decreases when $\tan\beta$ moves away from this point.

For the energies reachable by the LHC and a luminosity of $\unit[3]{ab^{-1}}$, we are able to exclude the moderate $\tan\beta$ region up to $\sim 1$ TeV, via $pp\to bbH/A \to bbtt$ for neutral Higgs bosons and via $pp\to tbH^\pm \to tbtb$ for charged Higgs bosons.
The search for low mass charged Higgses in the moderate $\tan\beta$ region is strongly affected by the pre-cuts on lepton transverse momentum and missing transverse energy and effectively keeps a small unprobed area, which we hope to cover with a slightly improved analysis.
This is indicated in Figure~\ref{fig:HA}.
Combining with $pp\to bbH/A \to bb \tau\tau$ and $pp\to tt H/A \to tttt$ (or $pp\to H/A \to tt$), a full coverage of $\tan\beta$ might be achievable for neutral Higgs searches, though a dedicated analysis is yet to be done for the latter.
As for the charged Higgs searches, the $pp\to H^\pm tb \to ttbb$ channel covers additionally the lower and higher $\tan\beta$ region up to more than \unit[2]{TeV}.
The discovery regions are more tightly constrained, at the HL-LHC the charged Higgs can be discovered in the associated channel for high and low $\tan\beta$ up to $\sim 2$ TeV.
% two associated channels are able to assist discovering neutral Higgs for masses up to \unit[500]{GeV} for moderate $\tan\beta$ and for
% charged Higgs up to $\sim 2$ TeV for high and low $\tan\beta$ and also probes the moderate $\tan\beta$ region up to $\sim \unit[1]{TeV}$.

At the \unit[100]{TeV} collider, a combination of $pp\to bb H \to bb \tau\tau$ and $pp \to bb H/A \to bb tt$ pushes the exclusion limit for the neutral Higgs searches up to $m_A \sim \unit[10]{TeV}$, except for the low $\tan\beta$ region.
For $pp \to bb H/A \to bb tt$, the low mass region has a worse sensitivity than the intermediate mass region, due to the pre-cuts on lepton transverse momentum and the missing transverse energy.
In Figure~\ref{fig:HA}, we also present the sensitivity reach of the $t_ht_l$ resonance search for reference (dashed red curves), with the interference effect between signal and background ignored. We however should not interpret it as the real reach for the neutral Higgs searches in the channel of $pp\to H/A \to tt$ at a \unit[100]{TeV} collider, since the interference effect can dramatically change the resonance structure.
Concurrently, $pp \to tb H^\pm \to tbtb$ pushes the exclusion limit for the charged Higgs searches up to $m_{H^\pm} \sim \unit[10]{TeV}$ with a full coverage of $\tan\beta$, with additional coverage up to $m_{H^\pm} \sim \unit[20]{TeV}$ for both high and low $\tan\beta$ regions.
Discovery of neutral and charged Higgs will be possible up to \unit[10]{TeV} and \unit[10--20]{TeV}, respectively.
In summary, with the channels under study, the HL-LHC and the \unit[100]{TeV} $pp$-collider has a potential to push the sensitivity reach for $H/A$ and $H^\pm$ from a scale of $\mathcal O(1)$ TeV up to a scale of $\mathcal O(10)$ TeV, respectively.

\section{Summary and Outlook}
\label{sec:dis}

The BSM Higgs sector is one of the most important physics targets at the LHC and at next-generation $pp$ colliders. In this article, we present a systematic study testing the MSSM Higgs sector in the decoupling limit, at \unit[14]{TeV} and \unit[100]{TeV} machines.
We propose that the ``top'' decay channels ($H/A \to tt$ and $H^\pm\to tb$) in associated Higgs productions, should play an essential role.
These channels are typically characterized by kinematics with highly boosted Higgs decay products, and large  forwardness/backwardness of the particles accompanying the Higgs production ($pp\to tt H/A$ is an exception, where $tt$ is less forward or backward).
Facilitated with a BDT method, the LHC will be able to cover the ``wedge'' region for neutral Higgs searches and hence test the Higgs sector up to $\unit[\mathcal O(1)]{TeV}$ for both, high and moderate $\tan\beta$ region.
We show that a future \unit[100]{TeV} $pp$-collider has a potential to push the sensitivity reach up to $\unit[\mathcal O(10)]{TeV}$.

The analyses pursued in this article represent a preliminary effort in this regard. A more complete study is definitely necessary.  Below are several directions that are interesting to explore in our view:
\begin{itemize}

\item Although, the channel $pp \to bb H/A \to bbtt$ enables us to cover the ``wedge'' or moderate $\tan\beta$ region for neutral Higgs searches in the decoupling limit, its sensitivity gets reduced below or around the threshold $m_H/m_A = 2 m_t$, where this channel either becomes kinematically forbidden at an on-shell level or yields soft decay products. However, we may take similar strategies, combining  the $bbH/A$ production with some other dominant decay modes in low $\tan\beta$ region, such as $H\to hh$ and $A\to hZ$. Then the unprobed ``wedge'' region might be covered by the channels $pp\to bb H \to bb hh$ and $pp\to bbA\to bbhZ$.

\item A realistic analysis for probing the low-$\tan\beta$ region in the neutral Higgs searches is still absent, largely because of the large interference effect between the channel $pp\to H/A \to tt$ signal and QCD $tt$ background.
To address this question, developing new strategies are definitely necessary and important.
One possible way out is to search for $pp\to tt H/A \to tttt$.
Although its cross section is relatively small, compared to the other channels, the strategies developed in this article might be of help.
Both topics are under exploration.
The results will be presented in a future article~\cite{in-prep}.

\item Although, the studies in this article are focused on the MSSM Higgs sector, it is straightforward to project their sensitivity reach to some other BSM scenarios, such as the 2HDM.
Additionally, the exploration can be generalized to exotic search channels.
Such channels might be switched on in a 2HDM model, but are generically suppressed in the MSSM~\cite{Coleppa:2014hxa, Coleppa:2014cca,Kling:2015uba,Li:2015lra}.
Dedicated analyses are certainly necessary.
We leave this exploration to future work.
\end{itemize}

\paragraph*{Note added:}
While this article was in finalization, the papers~\cite{Dev:2014yca, Craig:2015jba} appeared, which partially overlap with this one in evaluating the LHC sensitivities in searching for $pp\to bbH/A \to bbtt$ and $pp\to tbH^\pm \to tbtb$.  However, we notice a difference between the LHC sensitivities obtained in~\cite{Craig:2015jba} and in our analyses, which enable us to conclude that the LHC has a potential to fill up the well-known ``wedge'' (that is, the region with moderate $\tan\beta$) up to $\sim \unit[1]{TeV}$, with $\unit[3000]{fb^{-1}}$, via such channels.
In spite of this, we are focused more on testing an extended Higgs sector at a future \unit[100]{TeV} $pp$-collider, exploring the collider kinematics involved and developing its (BDT-based) search strategies, which are not covered in~\cite{Craig:2015jba}.

\section*{Acknowledgments}

We would like to thank T. Han, A. Ismail, A. V. Kotwal, I. Low, F. Maltoni, M. Mangano, M. Nojiri, J. Shu, S.-F. Su, Y.-J. Tu, L.-T. Wang, F. Yu, and H. Zhang for useful discussions.
The work is partly supported by the grant HKUST4/CRF/13G from the Research Grants Council (RGC) of Hong Kong.
Also, JH is supported by the grant of Institute for Advanced Study, HKUST;
TL and JS are supported by the grant at the HKUST; YL is supported by the the Hong Kong PhD Fellowship Scheme (HKPFS) issued by the Research Grants Council (RGC) of Hong Kong.
TL also would like to acknowledge the hospitality of the Kavli Institute for Theoretical Physics at UCSB, the Aspen Center for Physics (Simons Foundation) and the Institute for Advanced Study, HKUST, where part of this work was completed.

\newpage

\appendix

\section{Background Generation}

\begin{table}
\centering
\begin{tabular}{ll@{}r@{ }r@{ }r@{ }rrrr}
    \toprule
    Signal
  & Background
  & \multicolumn{3}{c}{Precut [GeV]}
  & \multicolumn{2}{c}{$\sigma$ [pb]}
  & \multicolumn{2}{c}{$\mathcal L_\text{gen}$ [$\unit{fb^{-1}}$]}
 \\ \cmidrule(r){3-5} \cmidrule(rl){6-7} \cmidrule(l){8-9}
  &
  & $p_T(t)$
  & $p_T(l)$
  & $E_T^\text{miss}$
  & \unit[100]{TeV}
  & \unit[14]{TeV}
  & \unit[100]{TeV}
  & \unit[14]{TeV}
 \\ \midrule
    \multirow{5}{*}{$H/Abb \to ttbb$}
  & \multirow{5}{*}{$tt \to bbjjl\nu$}
  & 2500
  &
  &
  & 0.0444
  &
  & 1570
 \\
  &
  & 1500
  &
  &
  & 0.505
  &
  & 384
 \\
  &
  & 300
  &
  &
  & 356
  &
  & 5.71
 \\
  &
  & 250
  &
  &
  &
  & 9.16
  &
  & 76.2
 \\
  &
  & 0
  &
  &
  & 7130
  & 166
  & 0.384
  & 10.8
 \\ \midrule
    \multirow{5}{*}{$H^\pm bb \to ttbb$}
  & \multirow{5}{*}{$tt \to bbjjl\nu$}
  & 2500
  &
  &
  & 0.0884
  &
  & 3453
 \\
  &
  & 1500
  &
  &
  & 0.894
  &
  & 255
 \\
  &
  & 300
  &
  &
  & 375
  &
  & 3.12
 \\
  &
  & 250
  &
  &
  &
  & 7.05
  &
  & 141
 \\
  &
  & 0
  &
  &
  & 7130
  & 166
  & 0.384
  & 10.8
 \\ \midrule
    \multirow{13}{*}{$H/Abb \to \tau\tau bb$}
  & \multirow{3}{*}{$bbZ/\gamma^* \to bb\tau\tau$}
  &
  & 700
  &
  & 0.000294
  &
  & 34000
\\
&
  &
  & 150
  &
  & 0.109
  &
  & 4600
\\
&
  &
  & 0
  &
  & 37.4
  &
  & 13.4
\\ \cmidrule{2-9}
  & \multirow{3}{*}{$ccZ/\gamma^* \to cc\tau\tau$}
  &
  & 700
  &
  & 0.000234
  &
  & 42700
\\
  &
  &
  & 150
  &
  & 0.0918
  &
  & 5450
\\
  &
  &
  & 0
  &
  & 65.8
  &
  & 7.60
 \\ \cmidrule{2-9}
  & \multirow{3}{*}{$tt \to bb\tau\tau\nu\nu$}
  &
  & 700
  &
  & 0.00187
  &
  & 4120
\\
  &
  &
  & 150
  &
  & 0.956
  &
  & 490.0
\\
  &
  &
  & 0
  &
  & 135.0
  &
  & 4.76%2.45
 \\ \cmidrule{2-9}
  & \multirow{3}{*}{$tt \to bbl\tau\nu\nu$}
  &
  & 700
  &
  & 0.00736
  &
  & 2590
\\
&
  &
  & 150
  &
  & 3.74
  &
  & 184.0
\\
&
  &
  & 0
  &
  & 600.0
  &
  & 2.13  %0.650
 \\ \midrule
    \multirow{7}[4]{*}{$H^\pm \to \tau\nu$}
  & \multirow{2}{*}{$tt\to bbjj\tau\nu$}
  &
  &
  & 300
  & 4.74
  &
  & 26.1
  \\
  &
  &
  &
  & 0
  & 428.0
  &
  &1.71
\\ \cmidrule{2-9}
&
  &
  &
  & 600
  & 0.0886 % 0.0886
  &
  & 1990 % 1989
 \\
  & $tt \to bb\tau\tau\nu\nu$
  &
  &
  & 300
  & 0.943
  &
  & 117.0
 \\
  &
  &
  &
  & 0
  & 135.0
  &
  & 4.76
 \\ \cmidrule{2-9}
  &
  &
  &
  & 600
  & 0.388 % 0.388
  &
  & 1540 % 1540
 \\
  & $tt \to bbl \tau\nu\nu$
  &
  &
  & 300
  & 4.32
  &
  & 36.8
 \\
  &
  &
  &
  & 0
  & 600.0
  &
  & 2.13
 \\ \bottomrule
\end{tabular}
\caption{Cross-sections and generated luminosities of the relevant background processes after precut.
All $tt$ backgrounds are matched up to two jets including $b$-jets.
The pre-cut $p_T(t)$ has been applied to both top quarks or the leading top quark and the leading non-top quark for neutral or charged Higgs, respectively.
The pre-cut $p_T(l)$ has been applied to all event leptons including $\tau$-leptons.
In our analysis we apply a k-factor of 1.2 to the LHC cross-sections~\cite{Cascioli:2013era}.
}
\label{tab:XsecBackground}
\end{table}

All background cross section are calculated with \software{MadGraph} and given to leading order.
In order to reduce the necessary background we apply different pre-cuts depending on Higgs mass.

For  the $pp \to bb H/A \to bb tt$ and  $pp \to tb H^\pm \to tb tb$ analyses the inclusive $tt$ background is generated with a pre-cut on the top transverse momentum of $p_T(t) = \unit[300, 1500, 2500]{GeV}$ for Higgs masses equal or larger to \unit[1000, 4000, 6000]{GeV}, respectively.
For the LHC analyses we have applied a pre-cut of \unit[250]{GeV} for masses equal or larger than \unit[1000]{GeV}.

In the $pp \to b b H/A \to b b \tau\tau $ analysis the irreducible background $b b Z/\gamma^* \to b b \tau\tau$ and the reducible backgrounds $c c Z/\gamma^* \to c c \tau\tau$, $t t \to b b \tau\tau  \nu \nu$ and $t t \to b b  l \tau \nu$ are considered.
We generate background samples with a pre-cut on the lepton transverse momentum of \unit[150]{GeV} for Higgs masses equal or larger than \unit[1000]{GeV}, and \unit[700]{GeV} for Higgs masses equal or larger than \unit[3000]{GeV}.

For the analysis $pp \to b t H^\pm \to b t \tau \nu$, the irreducible background is $t t \to b b j j \tau  \nu$ and the reducible backgrounds are $t t \to b b \tau \tau \nu \nu$ and $t t \to b b l \tau  \nu \nu$.
The background $t t \to b b j j \tau  \nu$ can be suppressed by a large transverse missing energy together with a requirement on a large transverse mass.
Hence we generate $t t \to b b j j \tau  \nu$ with transverse missing energy larger than \unit[300]{GeV} for Higgs masses larger than \unit[1]{TeV}, and for Higgs mass larger than \unit[3]{TeV}, the irreducible background can be well suppressed.
For $t t \to b b \tau \tau \nu \nu$ and $t t \to b b l \tau  \nu \nu$, the transverse missing energy is required to be equal or larger than \unit[300]{GeV} and \unit[600]{GeV} for Higgs masses larger than \unit[1]{TeV} and \unit[3]{TeV}, respectively.
All background cross-sections and the generated Luminosities are collected in Table~\ref{tab:XsecBackground}.
\section{BDT-based Analyses}

The analyses for $pp \to bb H/A \to bb tt$ and $pp \to tb H^\pm \to tb tb$ are based on BDT.

\subsection{Boosted Decision Trees (BDT)}
\label{sec:boosted decision trees}

A decision tree consists of a series of cuts classifying an event to be either signal or background.
Subsequent cuts of the decision tree are also applied to events previously classified as background, yielding shapes in the parameter space which approximate the signal region much better than the usual rectangular cuts.
The decision tree is trained on a training sample with truth level information; in a second stage it is applied to the testing sample without truth level information.
A decision tree can lead to a perfect separation between signal and background in the training sample.
However, this goes with an over-training effect where subsidiary cuts deal only with statistical effects of the training sample.
This can be seen by comparing the efficiency of the decision tree between the training sample and a test sample.
An enhancement to the decision tree dealing with this problem are boosted decision trees.
While the basic decision tree algorithm stays the same, it gets applied multiple times, on re-weighted samples.
The weighting factors are calculated from the region in parameter space where the decision tree has the lowest discriminating power.
BDT techniques have been introduced to high energy physics in~\cite{Roe:2004na} and have since then seen a large adoption especially in the experimental community, including~\cite{Ali:2011qf, ATLAS:2011tfa, CMS:2013fmk, TheATLAScollaboration:2013fja, Aad:2014aia}.
For a short introduction see~\cite{Coadou:2013lca}.
We make use of the \software{TMVA} package of the \software{ROOT} framework.
We are using the default AdaBoost algorithm and apply bagging, in order to further reduce over-training.
Our analysis code \software[0.1]{BoCA} is publicly available \cite{Boca}.

In order to classify the events of the training sample as signal or background we identify jets with their nearest quark.
Additionally, we define a set of observables according to which the BDT is trained.
In the testing and application phase the BDT is applied to a sample without truth level information and returns the signal likeliness of each element in this sample.
We combine multiple BDTs tailored to reconstruct single particles into a chain, which we apply to our events.
In a first step, we construct a simple bottom jet BDT tagger.
For boosted leptonic and hadronic top jets we develop two BDT taggers.
For non-boosted tops, on the other hand, we first reconstruct hadronic and leptonic $W$-bosons, subsequently we combine these $W$-bosons with a $b$-jet.
In the next step we reconstruct neutral or charged Higgs from two tops or one $b$-jet and one top, respectively.
Additionally, we search for the bottom-fusion jet pairs with large difference in rapidity.
Finally, we combine the reconstructed Higgs and the bottom-fusion pair and distinguish signal and background events with a last BDT analysis.

\begin{enumerate}
\item Bottom Tagger

In order to distinguish bottom jets from light jets we make use of the finite lifetime of the intermediate bottom mesons.
A bottom jet has multiple displaced vertices with a larger invariant mass and energy fraction compared to light jets.
Therefore, we train the bottom jet BDT on
\begin{description}
 \item[Vertex Mass] invariant mass of tracks with displaced vertex $m_V$
 \item[Multiplicity] number of tracks with displacement $N_t$
 \item[Displacement] cylindrical distance between the primary vertex and the secondary vertices $\Delta r$
 \item[Energy fraction] between tracks with and without displacement $r_E$
 \item[Radius] of the jet associated with a secondary vertex $\Delta R_j$
 \item[Spread] concentration of transverse momentum towards the center of the jet \\ $\sum_{j_c\in j}\frac{\Delta R(j_c,j) p_T^{j_c}}{p_T^j \Delta R_j}$
\end{description}

\item Top Tagger

For highly boosted tops we identify the top jets solely based on the global jet variables, the information originating in the displacement of the $b$-mesons and the possible presence of a lepton inside the jet cone.
Additionally to the variables used for the bottom tagger we use

\begin{description}
\item[Mass] of the jet
\item[Hardness] of the leading lepton inside the jet cone in the case of leptonically decaying tops.
For hadronically decaying tops we veto on hard leptons inside the jet cone.
\end{description}

\begin{figure}
\begin{subfigure}{.45\textwidth}
\includegraphics[width=\textwidth]{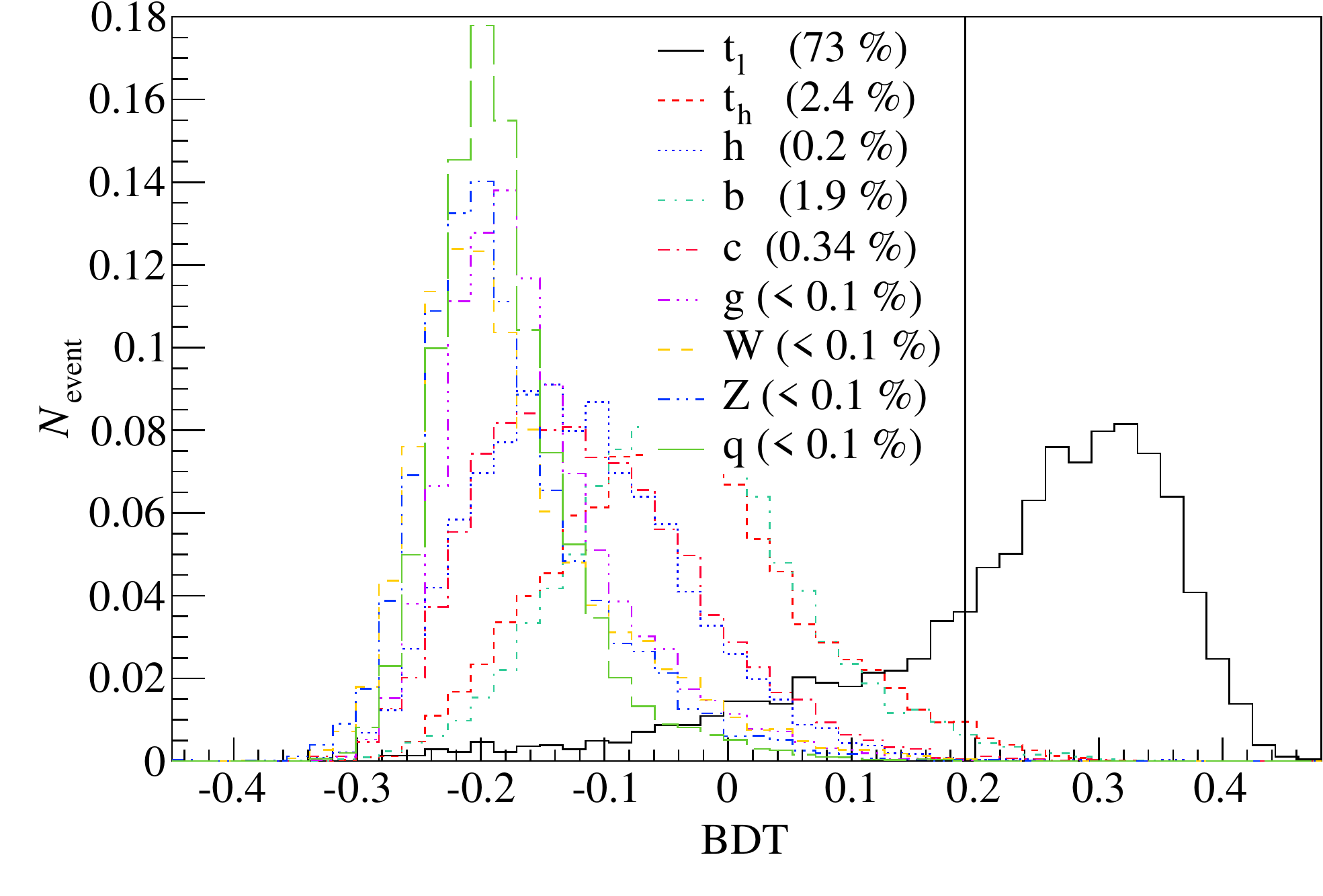}
\caption{$\unit[500]{GeV} < p_T^j < \unit[1000]{GeV}$}
\label{fig:leptonic top low pt}
\end{subfigure}
\hfill
\begin{subfigure}{.45\textwidth}
\includegraphics[width=\textwidth]{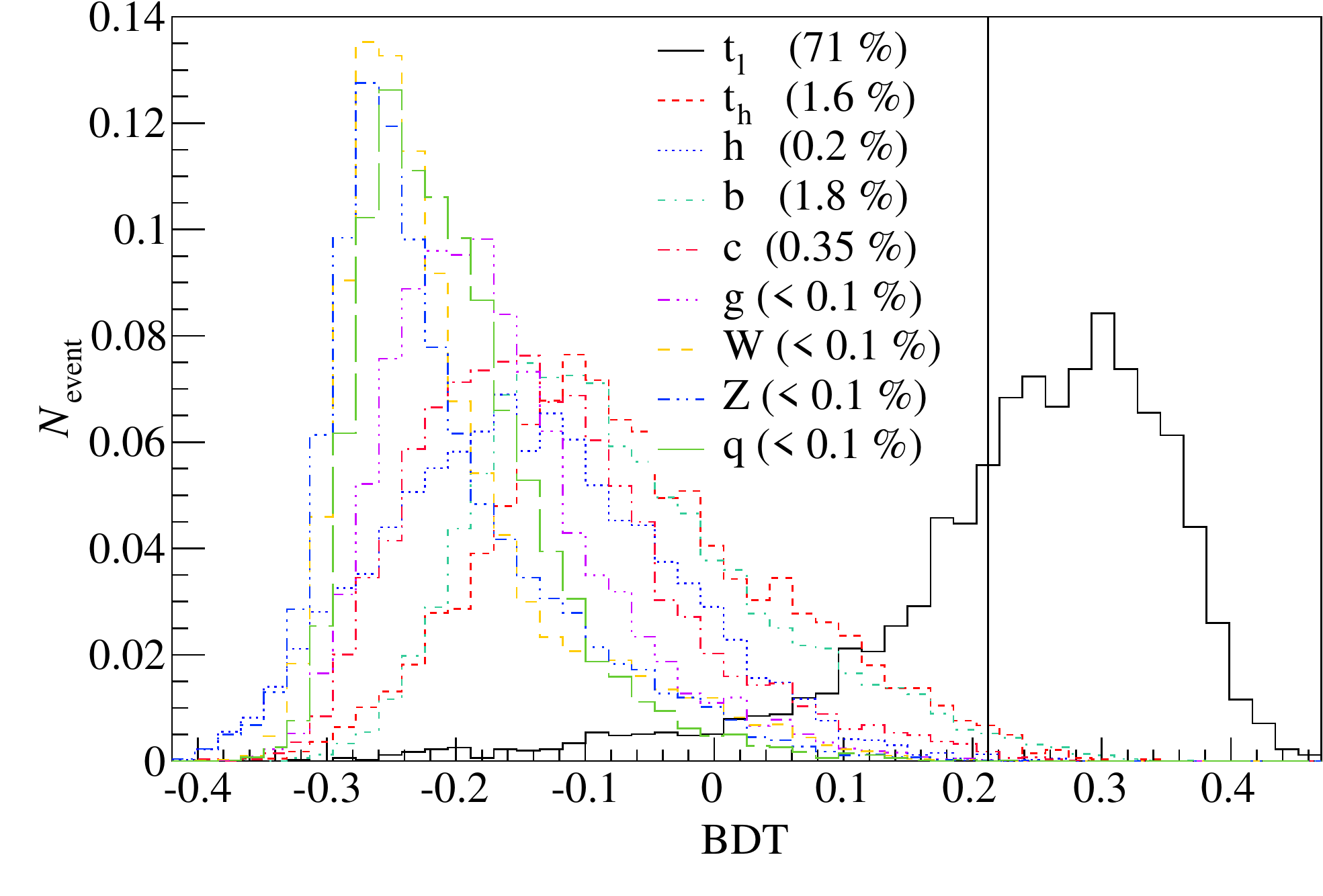}
\caption{$\unit[1000]{GeV} < p_T^j < \unit[1500]{GeV}$}
\label{fig:leptonic top high pt}
\end{subfigure}
\caption{%
BDT distributions for leptonic top.
We present signal (solid) and background (dotted) for two different transverse momenta regions.
The low $p_T$ region comprises of $\unit[500]{GeV} < p^j_T < \unit[1000]{GeV}$ and is presented in Figure~\subref{fig:leptonic top low pt}.
The high $p_T$ region comprises of $p^j_T > \unit[1000]{GeV}$ and is presented in Figure~\subref{fig:leptonic top high pt}.
All samples have a size of 10000. For their definition, see the caption of Fig.~\ref{fig:hadronic top tagger}.
}
\label{fig:Top_lep}
\end{figure}

\begin{figure}
\begin{subfigure}{0.48\textwidth}
\includegraphics[width=\textwidth]{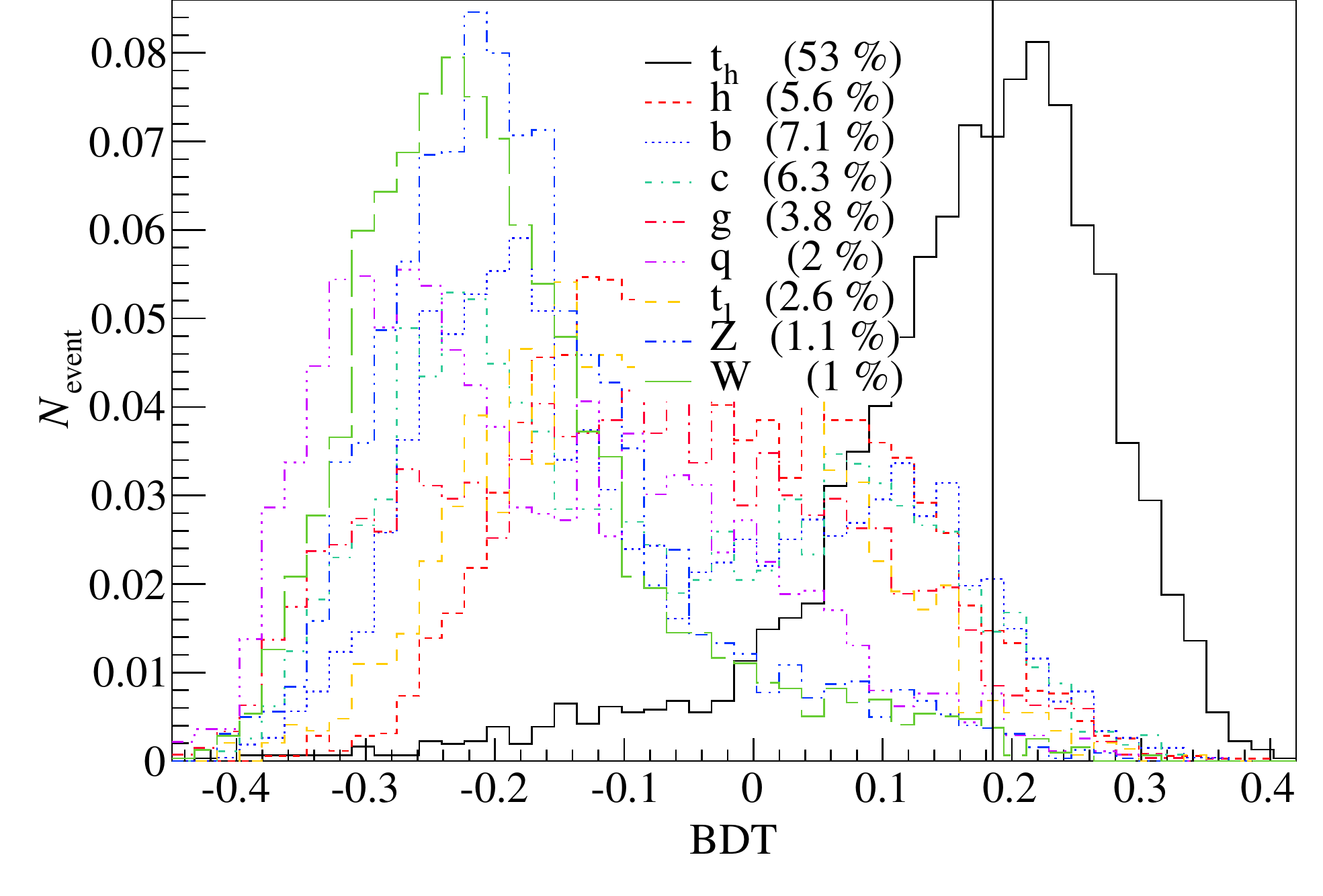}
\caption{$\unit[700]{GeV} < p_T^j < \unit[1000]{GeV}$}
\end{subfigure}
\hfill
\begin{subfigure}{0.48\textwidth}
\includegraphics[width=\textwidth]{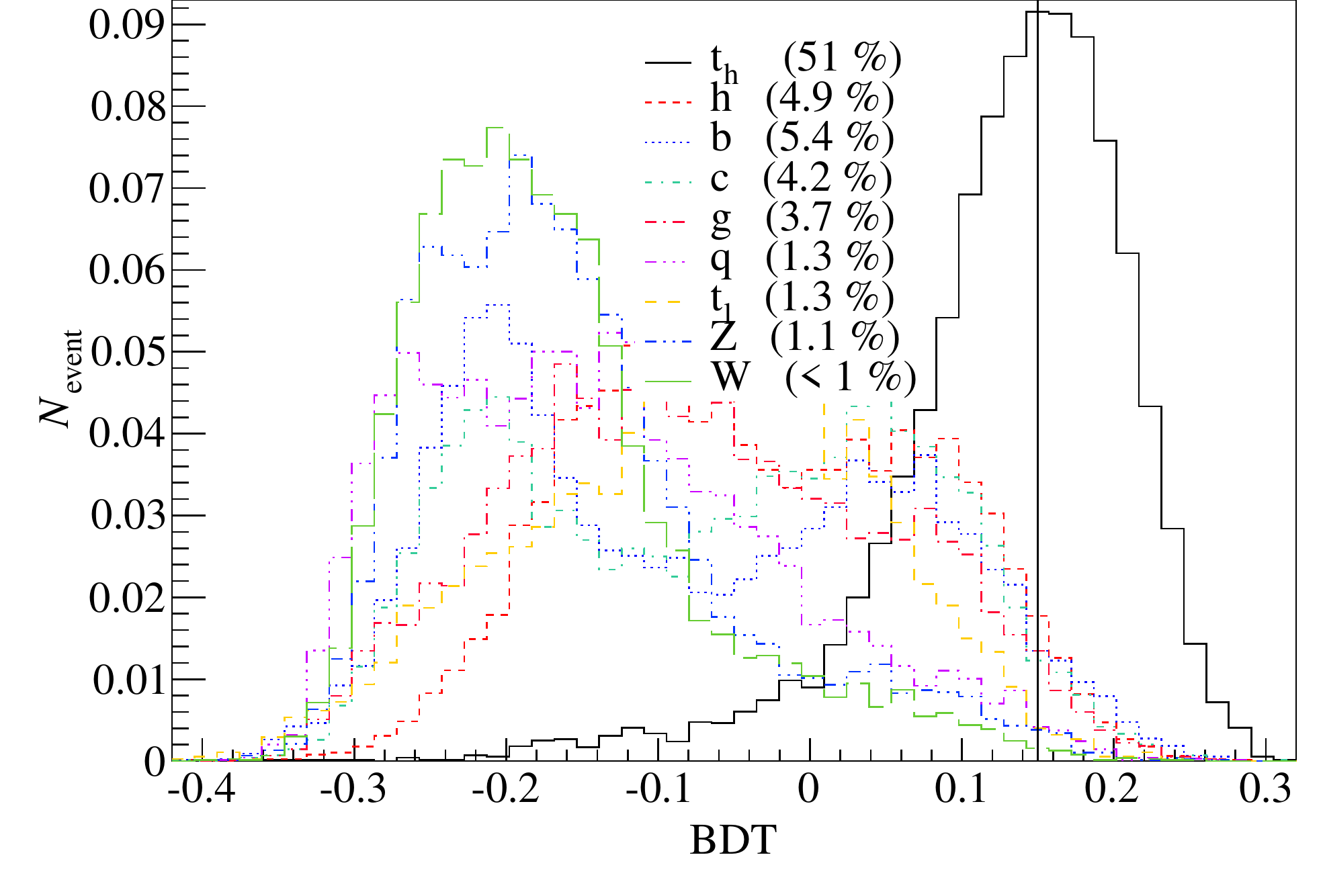}
\caption{$\unit[1000]{GeV} < p_T^j < \unit[1500]{GeV}$}
\end{subfigure}
\caption{BDT distribution of the hadronic top tagger.
We present signal (solid) and background (interrupted) for two different ranges of transverse momenta.
The particles have been generated with a transverse momentum pre-cut of \unit[500 and 1000]{GeV}, respectively.
Additionally the jets are required to fall into the mass-windows mentioned in the figure description.
All samples have a size of 10000. For their definition, see the caption of Fig.~\ref{fig:hadronic top tagger}.
}
\label{fig:hadronic top bdt}
\end{figure}

Figure~\ref{fig:Top_lep} shows the BDT-response of the leptonic top tagger to different jets with the same boostness.
The light-, $W$- and $Z$-jet fake rate is mostly suppressed by the displacement information.
The track multiplicity and vertex mass reduces the fake rate for $h$-jets as the main decay channel into $b$-quarks yields more secondary vertices compared to leptonic top jet.
The jet mass distribution of the heavy quarks are centered around the respective quark and boson masses, which reduces the fake rates of theses jets further.
On the other hand, the jet masses of the light quarks and gluons form a very broad slope, maximized at zero, which makes it hard to suppress the fake rate of these jets with the jet mass information.

In Figure~\ref{fig:hadronic top bdt} we present the BDT-response of the hadronic top tagger to different jets with the same boostness.
Also in this case the main discriminator against light-, $W$- and  $Z$-jets is the presence of secondary vertices.
Higgs jets are suppressed by the energy fraction between tracks with and without secondary vertices.
The presence of an hard lepton suppresses the leptonic top.
In the case of less boosted hadronic top jets we make use of the information provided by its decay products.
Hence, we probe if the jet can be re-clustered into two sub-jets to resolve the $b$ and $W$ component or into three sub-jets to resolve additionally the decay products of the $W$.

\item Top reconstruction
\label{item:top reconstruction}

For non-boosted top jets we have to assume that the $W$ and the $b$ are located in two or three different jets.
Hence, we perform a $W$ reconstruction whenever necessary, which combines two (sub-)jets and probes on
\begin{description}
 \item[Invariant mass] of the reconstructed object
 \item[Momentum-space position] of the reconstructed object ($p_T$, $\eta$, $\phi$)
 \item[Angular difference] between both elements of the reconstructed object ($\Delta \eta$, $\Delta \phi$, $\Delta R$)
 \item[Momentum difference] $\Delta p_T$
 \item[BDT response] of the preceding reconstruction step; In the case of hadronic $W$-reconstruction this  implies a veto on $b$-jets.
\end{description}
In a second step we combine the the reconstructed or tagged $W$'s with a $b$-jet to a top and train the top BDT on the same variables as in the $W$-BDT.

Finally, we have to take the case of non-boosted leptonic top into account.
We consider the hardest lepton as possibly originating from a $W$ decay.
In order to reconstruct these leptonically decaying $W$-bosons, we reconstruct the neutrino four momentum from the lepton momentum and missing energy.
Whenever the appearing quadratic equation has complex solutions we iteratively move the missing energy vector towards the lepton until the solution becomes real.
Of these two solution we train the BDT on the one that coincides better with the truth level neutrino.

\item Heavy Higgs reconstruction

For the Higgs reconstruction we combine two tops and one top with one bottom for neutral and charged Higgs, respectively.
We train the BDT on the same variables as in the top reconstruction (cf. Step~\ref{item:top reconstruction}).

\item Bottom Fusion BDT

Additionally we train a BDT on the bottom fusion pair accompanying the Higgs.
The most important discriminating variables in this case are the large rapidity difference between these two jets and their $b$- (top-) likeliness.
We demonstrate the performance of this tagger on a vector boson fusion samples in Figure~\ref{fig:bottom fusion tagger}.

\item Event BDT

In the last step we combine the reconstructed heavy Higgs with the bottom fusion pair and global event observables.
The global variables contain
\begin{description}
 \item[\boldmath $H_T$] of the event
 \item[Jet number]
 \item[Lepton number]
 \item[Bottom BDT] response of the remaining jets
 \item[Hardness] of the remaining jets and leptons
\end{description}
\end{enumerate}
% The result of this final BDT is demonstrated in Figure~\ref{fig:Event BDT}

% \subsection{Numerical Results}

\begin{figure}
\begin{subfigure}{0.48\textwidth}
\includegraphics[width=\textwidth]{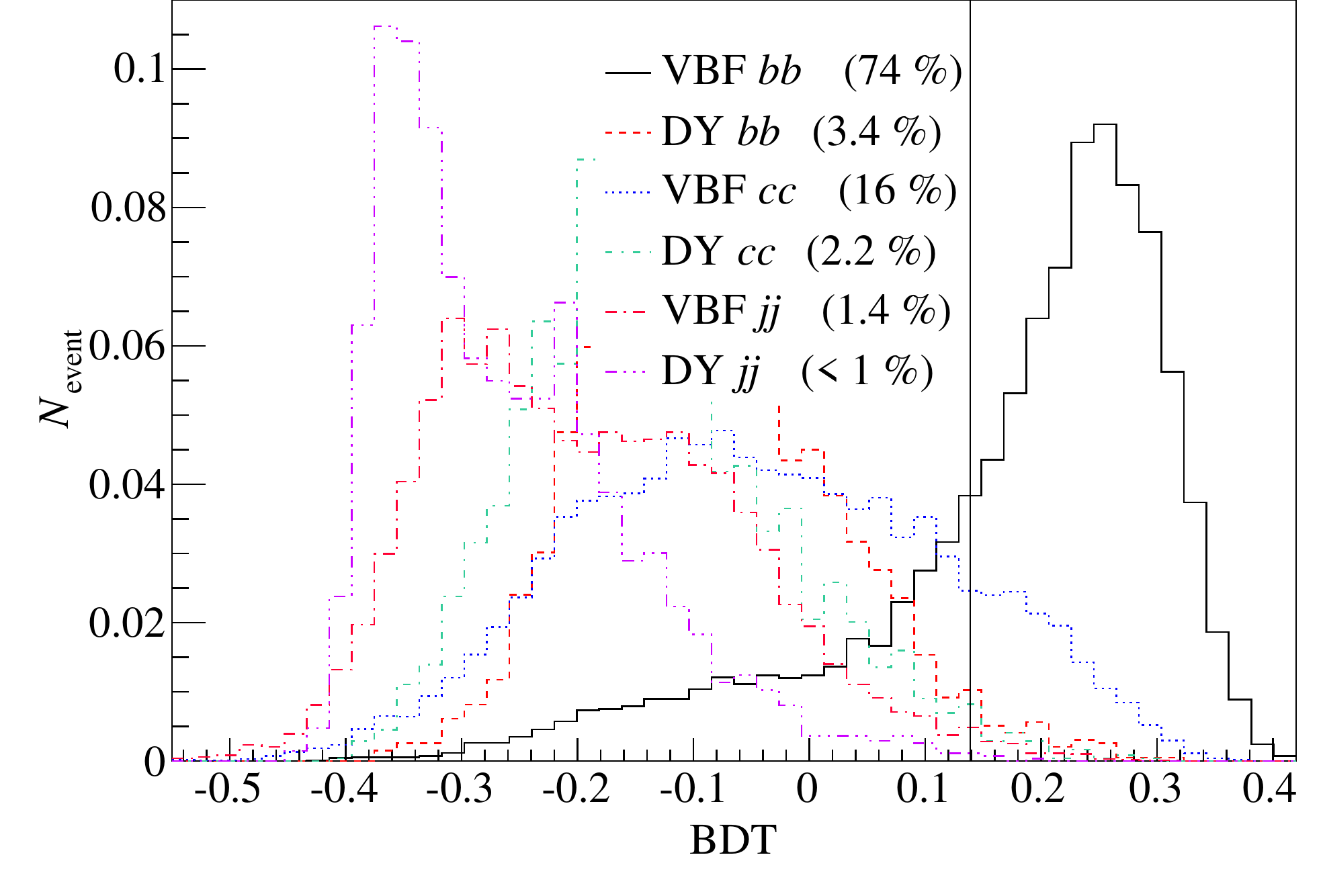}
\caption{Bottom fusion pair BDT.}
\label{fig:bottom fusion tagger}
\end{subfigure}
\hfill
\begin{subfigure}{0.48\textwidth}
\includegraphics[width=\textwidth]{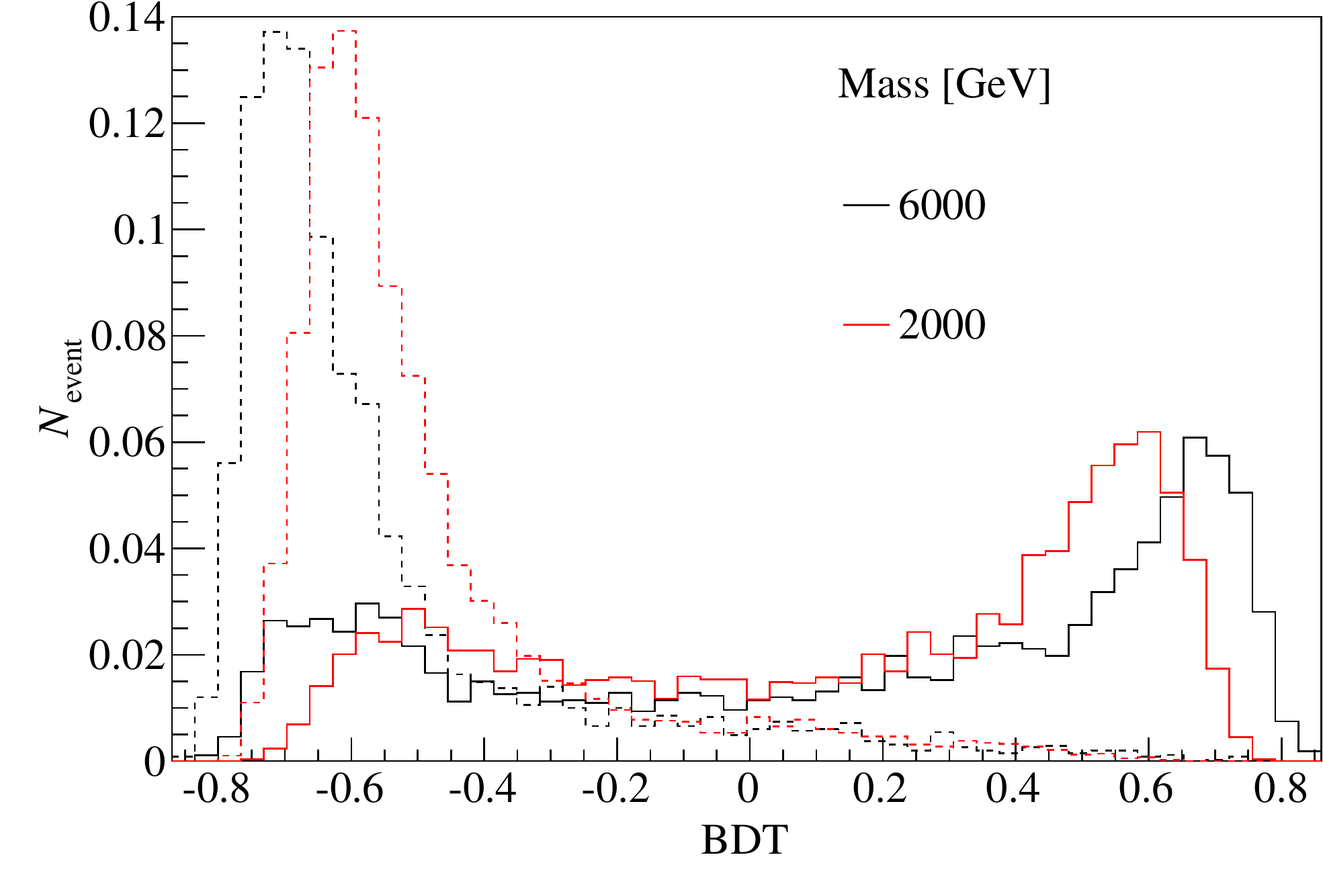}
\caption{$H/Abb$ event BDT.}
\label{fig:event bdt tagger}
\end{subfigure}
\caption{%
BDT distributions for the bottom fusion pair BDT~\subref{fig:bottom fusion tagger} and the final event BDT for associated production of neutral Higgs~\subref{fig:event bdt tagger} at a 100 TeV $pp$ collider.
The bottom fusion pair BDT~\subref{fig:bottom fusion tagger} is tested with $bb$, $cc$ and $jj$ vector boson fusion and Drell-Yan samples.
The large overlap between the bottom and the charm sample is due to the limited tracker coverage of $|\eta|<3.5$.
For the event BDT~\subref{fig:event bdt tagger} we show two examples of signal (solid) and background (dotted).
}\label{fig:Event BDT}
\end{figure}

We illustrate the response of the event BDT for the case of neutral Higgs in Figure~\ref{fig:event bdt tagger}.
% The numerical results are summarized in Table~\ref{tab:BDT}.

\section{Cut-based Analyses}

We analyze the processes $pp \to b b H/A \to b b \tau_h \tau_l $ and $pp \to b t H^\pm \to b t \tau_h \nu$ at 100 TeV, with a cut based approach.

\subsection{Neutral Higgs}

\begin{table}
\centering
\tabcolsep=0.5em
\begin{tabular}{lrrrrrr}
    \toprule
  & \multicolumn{2}{c}{Signal}
  & \multicolumn{4}{c}{Background}
 \\ \cmidrule(rl){2-3} \cmidrule(l){4-7}
  & $b b H \to b b \tau\tau$
  & $b b A \to b b \tau \tau$
  & $b b Z/\gamma^{*} \to b b \tau\tau$
  & $c c Z/\gamma^{*} \to c c \tau \tau$
  & \multicolumn{2}{c}{$t t$  }
 \\ \cmidrule(l){6-7}
  &
  &
  &
  &
  & $b b \tau\tau \nu\nu$
  & $b b  l \tau \nu\nu$
\\ \midrule
    $N_\text{gen}$
  & 10000
  & 10000
  & 10000
  & 10000
  & 7686
  & 19043
 \\ \cmidrule(r){1-1} \cmidrule(rl){2-3} \cmidrule(l){4-7}
 $N_\text{cut 1}$
  & 736
  & 756
  & 403
  & 451
  & 195
  & 5699
 \\ $N_\text{cut 2}$
  & 192
  & 185
  & 64
  & 58
  & 124
  & 2899
 \\ $N_\text{cut 3}$
  & 123
  & 114
  & 31
  & 26
  & 39
  & 470
 \\ $N_\text{cut 4}$
  & 16
  & 10
  & 11
  & 4
  & 4
  & 91
 \\ $N_\text{cut 5}$
  & 10
  & 6
  & 2
  & 0
  & 0
  & 7
\\ \cmidrule(r){1-1} \cmidrule(rl){2-3} \cmidrule(l){4-7}
  $f_s$
  & 0.460
  & 0.460
  & 0.088
  & 0.07
  & 0.73
  & 1.16
 \\ \bottomrule
\end{tabular}
\caption{%
Cut flows for $pp \to b b H/A \to b b \tau\tau$ signal and background with $m_H/m_A = \unit[3]{TeV}$.
The scale factor $f_s$ is calculated for a luminosity of $\unit[3000]{fb^{-1}}$ and total cross section $\sigma = \unit[1.53]{fb}$,  corresponding to $\tan \beta = 10$.
The signal and background samples are generated with $\tau$ decaying inclusively.
During the generation of the background sample, a pre-cut of $p_T > \unit[700]{GeV}$ is applied to both leptons and the $\tau$-leptons.
}
\label{tab:Cut_flow_neutral}
\end{table}

For the semi-leptonic channel $H/A \to \tau_l\tau_h$ with a large Higgs masses, the signal contains one hard lepton and one hard jets possibly tagged as $\tau$-jet.
We present the analysis for $m_H/m_A = \unit[3]{TeV}$ for illustration.
The cuts applied in our analysis are
\begin{description}
\item[Cut 1] One lepton with $p^l_T > \unit[700]{GeV}$ and veto on a second lepton with $p^l_T > \unit[25]{GeV}$
\item[Cut 2] Exactly one oppositely charged ($Q(\tau)=-Q(l)$) jet with $\tau$-tag and $p^{\tau}_T > \unit[700]{GeV}$
\item[Cut 3] Transverse mass $m_T(l, E_T^\text{miss}) < \unit[160]{GeV}$
\item[Cut 4] Exactly two b-tag jets with $p^b_T > \unit[40]{GeV}$
\item[Cut 5] $\Delta\eta_{bb} > 2.5$
\end{description}
Here $m_T(l, E_T^\text{miss})$ is transverse mass defined as
\begin{align}
    m_T(l, E_T^\text{miss})
  = \sqrt{2p^l_T E_T^\text{miss} ( 1 - \cos \phi_{l, \text{miss}})}
\ .
\label{transverse mass}
\end{align}
where $\phi_{l, \text{miss}}$ is the azimuthal angle between lepton and the missing transverse momentum.
The cut flows for signal and background are presented in Table~\ref{tab:Cut_flow_neutral}. The resulting significance for exclusion is $\unit[1.86]{\sigma}$ for a luminosity of $\unit[3000]{fb^{-1}}$.
The resulting exclusion region at a \unit[100]{TeV} collider is shown in Figure~\ref{fig:HA}.

\subsection{Charged Higgs}

The kinematics for the channel $pp\to tb H^\pm \to t b \tau_h \nu $ with a large Higgs mass is characterized by large missing energy, hard $\tau_\text{had}$ and large $m_T(\tau_\text{had}, E_T^\text{miss})$.
These features suppress the irreducible and the multi-jet background, particularly in the large Higgs mass domain.

\begin{table}
\centering
\begin{tabular}{lrrr}
    \toprule
  & \multicolumn{1}{c}{Signal}
  & \multicolumn{2}{c}{Background}
 \\ \cmidrule(rl){2-2} \cmidrule(l){3-4}
  & $tbH^\pm\to tb\tau \nu$
  & $t t \to b b \tau \tau \nu\nu$
  & $t t \to b b l \tau \nu \nu$
 \\ \midrule
    $N_\text{gen}$
  & 10000
  & 176203
  & 598242
 \\ \cmidrule(r){1-1} \cmidrule(rl){2-2} \cmidrule(l){3-4}
    $N_\text{cut 1}$
  & 491
  & 2334
  & 1874
 \\ $N_\text{cut 2}$
  & 118
  & 20
  & 8
 \\ \cmidrule(r){1-1} \cmidrule(rl){2-2} \cmidrule(l){3-4}
    $f_s$
  & 0.152
  & 1.51
  & 1.95
 \\ \bottomrule
\end{tabular}
\caption{Cut flows for $pp \to t b H^\pm \to t b \tau\nu$ signal and background with $m_{H^\pm} = \unit[3]{TeV}$, where $f_s$ is the scale factor for a luminosity of $\unit[3000]{fb^{-1}}$ and total cross section $\sigma = \unit[0.507]{fb}$, corresponding to $\tan \beta = 10$.
The signal and background samples are generated with $\tau$ decaying inclusively.
For $m_{H^\pm}=3$ TeV, the irreducible background $t t \to b b j j \tau  \nu$ and multi-jet background are well suppressed. }
\label{tab:Cut_flow_charged}
\end{table}

We present the analysis for $m_{H^\pm}=\unit[3]{TeV}$ for illustration. Below are the analyses cuts applied:
\begin{description}
\item[Cut 1] $\ge 4$ jets, with exactly two $b$-tagged, $\Delta\eta_{bb} \ge 2.5$, and veto of leptons with $p_T > \unit[25]{GeV}$.
\item[Cut 2] exactly one hard ($p_T > \unit[650]{GeV}$) $\tau$-tagged jet,  $E_T^\text{miss} > \unit[650]{GeV}$ and transverse mass $m_T(\tau, E_T^\text{miss}) > \unit[1300]{GeV}$.
\end{description}
We present the cut flows for signal and background in Table~\ref{tab:Cut_flow_charged}. The significance for exclusion is $\unit[2.25]{\sigma}$ for a luminosity of $\unit[3000]{fb^{-1}}$.
The cuts on $E_T^\text{miss}$, $p_T(\tau_\text{had})$ and $m_T$ are modified for optimizing the analyses, as $m_{H^\pm}$ varies.
The resulting exclusion region at a \unit[100]{TeV} collider is shown in Figure~\ref{fig:HA}.

\bibliographystyle{JHEP}
\bibliography{references}

\end{document}

%% file: figs/FlowChartBoosted.pgf
% \begin{tikzpicture}[-{Latex[length=.8ex]}, inner sep=.5ex,node distance=2.5ex and 2.5em]
\begin{tikzpicture}[->, inner sep=.5ex,node distance=2.5ex and 2.5em]
\tikzset{box/.style = {rounded corners=1.5ex, draw,inner sep = 1.3ex}}

\node(eflow){EFlow\strut};
% \node(l)[below = of eflow]{$l$};
% \node(etmiss)[below = of l]{$E_T^{\rm miss}$};
\node(del) [fit = (eflow), box]{};
\node(delphes) [above = 1ex of del]{\tt Delphes\strut};

\node (clustering) [right = of eflow] {clustering\strut};
\node(j)[below = 1ex of clustering]{$j$};
\draw (eflow) -- (j);

\node(b)[right = of clustering]{$b$};
\draw[dashed] (j) -> (b);
\node (space) [left = 0em of b]{};

\node (fj) [box , fit =(j) (clustering)]{};
\node [above = 1ex of fj] {\tt FastJet\strut} ;

\node (th)[ below = of b]{$t_h$};
\draw[dashed] (j)--(th);

\node(tl) [below = of th ]{$t_l$};
% \draw[dashed] (l) -> (tl);
\draw [dashed](j)-- node[below left]{($l$)}(tl);
% \draw (etmiss) -> (tl);

\node (pair)[right = of b]{Pair\strut};
\draw (b)--(pair);

\node (h)[right = of th]{$H/A$};
\draw (th)--(h);
\draw (tl)--(h);

\node (sig)[ right = of h]{Signature\strut};
\draw (h)--(sig);
\draw (pair)--(sig);

\node (event)[below = of sig]{Event};
\draw (sig)--(event);

\node (ana) [box, fit =(space)(b)(tl)(th)(pair)(h)(sig)] {};
\node [above = 1ex of ana]{Analysis\strut};
\end{tikzpicture}

%% file: figs/FlowChart.pgf
% \begin{tikzpicture}[-{Latex[length=.8ex]}, inner sep=.5ex,node distance=2.5ex and 2.5em]
\begin{tikzpicture}[->, inner sep=.5ex,node distance=2.5ex and 2.5em]
\tikzset{box/.style = {rounded corners=1.5ex, draw,inner sep = 1.3ex}}

\node(eflow){EFlow};
\node(l)[below = of eflow]{$l$\strut};
\node(etmiss)[below = of l]{$E_T^{\rm miss}$\strut};
\node(del) [fit = (l)(etmiss)(eflow), box]{};
\node(delphes) [above = 1ex of del]{\tt Delphes\strut};

\node (clustering) [right = of eflow]{clustering};
\node(j)[below = 1ex of clustering]{$j$\strut};
\draw (eflow) -- (j);

\node(b)[ right = of clustering]{$b$};
\draw[dashed](j) -> (b);
\node (space) [left = 0em of b]{};

\node (fj) [box , fit =(j) (clustering)]{};
\node [above = 1ex of fj] {\tt FastJet\strut} ;

\node (Wh)[ below = of b]{$W_h$\strut};
% \draw[dotted] (b)--(Wh);
\draw (j)--(Wh);

\node(Wl) [below = of Wh ]{$W_l$\strut};
\draw (l) -> (Wl);
\draw (etmiss) -> (Wl);

\node (phantom) [right = of b]{\phantom{ M }};
\node (pair)[right = of phantom]{Pair\strut};
\draw (b)--(pair);

\node (th)[right = of Wh]{$t_h$};
\draw (b)--(th);
\draw (Wh)--(th);

\node (tl)[right = of Wl]{$t_l$};
\draw (b)--(tl);
\draw (Wl)--(tl);

\node (h)[ right = of th]{$H/A$};
\draw (th)--(h);
\draw (tl)--(h);

\node (sig)[right =  of h]{Signature};
\draw (h)--(sig);
\draw (pair)--(sig);

\node (event)[below = of sig]{Event};
\draw (sig)--(event);

\node (ana) [box, fit =(space)(b)(Wl)(Wh)(tl)(th)(pair)(h)(sig)] {};
\node [above = 1ex of ana]{Analysis\strut};
\end{tikzpicture}